\documentclass{article}
\usepackage{graphicx} 
\usepackage{amsmath}
\usepackage{hyperref}
\usepackage{authblk}
\usepackage{tcolorbox}
\usepackage{enumitem}
\usepackage{cite}

\title{Impermanent loss and Loss-vs-Rebalancing II}

\author{
  Abe Alexander\thanks{abealexander@outlook.com} \and
  Guillaume Lambert\thanks{glambert@panoptic.xyz} \and
  Lars Fritz\thanks{lsfritz@proton.me}
}

\begin{document}

\maketitle
\begin{abstract}
This paper examines the relationship between impermanent loss (IL) and loss-versus-rebalancing (LVR) in automated market makers (AMMs). Our main focus is on statistical properties, the impact of fees, the role of block times, and, related to the latter, the continuous time limit. We find there are three relevant regimes: (i) very short times where LVR and IL are identical; (ii) intermediate time where LVR and IL show distinct distribution functions but are connected via the central limit theorem exhibiting the same expectation value; (iii) long time behavior where both the distribution functions and averages are distinct. Subsequently, we study how fees change this dynamics with a special focus on competing time scales like block times and 'arbitrage times'. 
\end{abstract}

\section{Introduction}

Automated Market Makers (AMMs) are a cornerstone innovation in decentralized finance (DeFi), enabling digital asset trading without relying on a traditional order book. Instead, AMMs leverage liquidity pools and pricing algorithms to execute trades. Liquidity providers (LPers) supply the assets that constitute these pools but face risks such as impermanent loss (IL) or loss-versus-rebalancing (LVR). These risks arise when the value of deposited assets changes compared to holding the assets individually or continuously rebalancing a portfolio.

This paper investigates the interplay between IL and LVR in AMMs, examining their statistical properties and the influence of fees on AMM performance. We show that for infinitesimal price movements, IL and LVR exhibit identical behavior, despite differing in interpretation. However, over extended time ranges, their distributions and dynamics diverge considerably. This divergence becomes even more pronounced when fees are introduce: aside from introducung a novel time scale we also find that fees impact the two metrics differently.

The paper is structured as follows: In Sec.~\ref{sec:setup} with discuss the mathematics of the simplest form of an AMM, the constant product market maker. Furthermore, we introduce the concept of Brownian motion and geometric Brownian motion to discuss the price dynamics of an asset pair. We argue that for short and intermediate times, the difference between both is mostly irrelevant, whereas for longer time scales it becomes pronounced. The reason we also discuss Brownian motion is that it allows for simpler analytic discussions. In Sec.~\ref{sec:ilvslvr} we introduce the concepts of IL and LVR. We then move to the arbitrage dynamics of an AMM that interacts with an infinite liquidity source in Sec.~\ref{sec:arbnofees} in the absence of fees. This discussion is carried out in three different time regimes, called short, intermediate, and long. An analytic discussion is relegated to App.~\ref{app:illvr} and a discussion of the connection between IL and LVR via the central limit theorem in App.~\ref{app:clt}. Afterwards, in Sec.~\ref{sec:illvrfee} we analyze arbitrage dynamics in the presence of fees and show how fees introduce a novel time scale, the average arbitrage time. We end with a conclusion in Sec.~\ref{sec:conc}.

Throughout the paper, we combine analytical arguments with numerical simulations to provide a comprehensive understanding of these critical AMM metrics and their implications for decentralized finance. We choose intuition over mathematical rigor and take a mostly explicit and self-contained approach. We hope that this makes the article accessible to a wide set of readers with a variety of backgrounds. Many of the results in this paper provide a more pedestrian rederivation of mathematically more rigorous prior discussion, especially those presented in Refs.~\cite{evans2021optimalfeesgeometricmean,milionis2024automated,milionis2023automated,fritsch2024measuringarbitragelossesprofitability}. However, according to our knowledge, the role of distribution functions has not been discussed and presented elsewhere in detail (There is a very recent discussion in Ref.~\cite{fritsch2024mevcapturetimeadvantagedarbitrage} in which the distribution of a quantity appears that resembles the distribution function of IL that we find). We believe that this work gives an important new angle on the relation between IL and LVR. From a pratical point of view it shows that suppressing LVR does not protect the LPer from incurring huge losses due to IL or divergence loss. However, it also shows that protecting LPers from LVR makes a very positive contribution to also mitigate the most probable forms of IL. This paper is a more detailed and expanded sequel to a recent paper, Ref.~\cite{AlexanderFritz3}.

\noindent{\bf{Related literature:}}

 AMMs can be traced back to \cite{hanson2007logarithmic} and \cite{othman2013practical} with early implementations discussed in \cite{lehar2021decentralized}, \cite{capponi2021adoption}, and \cite{hasbrouck2022need}. Details of implementation are described in \cite{Adams20} and \cite{Adams21} as well as in a very recent textbook \cite{ottina2023automated}.

Discussions of fees and how use them to mitigate LPers' losses appear in several places: Uniswap v3 (\cite{Adams21}) addresses this problem by letting liquidity providers choose between different static fee tiers. Other automated market makers have implemented dynamic fees on individual pools, including Trader Joe v2.1 (\cite{mountainfarmer22joe}), Curve v2 (\cite{egorov21curvev2}) and Mooniswap (\cite{bukov20mooniswap}), Algebra (\cite{Volosnikov}), as well as \cite{Nezlobin2023}. Some of the general properties of toxic flow and loss versus rebalancing and fees have been discussed in Refs.~\cite{evans2021optimalfeesgeometricmean,Fritsch_2021,Faycal1,Faycal2,Faycal3,milionis2024automated,canidio2024arbitrageursprofitslvrsandwich,fritsch2024mevcapturetimeadvantagedarbitrage,fritsch2024measuringarbitragelossesprofitability,AElsts2024,AlexanderFritz1,AlexanderFritz2,AlexanderFritz3}.

\section{The setup}\label{sec:setup}

In this section we discuss two vital technical points: the concept and properties of a constant function market maker AMM as well as the concept of price modeling by means of (geometric) Brownian motion. The discussion is intended to be basic and self-contained. We encourage more experienced readers to skip this section.

\subsection{The automated market maker}

A constant function market maker (CFMM) with the formula \(xy - L^2 = 0\) describes the most basic automated market maker (AMM) model where \(x\) and \(y\) represent the quantities of two different tokens in a liquidity pool, and \(L\) is a constant that characterizes the pool's total liquidity, a measure for its resistance to price changes. In this model, the product of the quantities of the two tokens remains constant:

\[
xy = L^2\;.
\]

This ensures that any trade which increases one token's amount \(x\) must decrease the other token's amount \(y\) and vice versa. The price of each token depends inversely on their respective quantities. Specifically, the price of token \(x\) in terms of token \(y\) is given by the so-called spot price

\[
p = \frac{y}{x}\;,
\]

henceforth simply referred to as price. There is a relation between the token amounts and the price $p$ according to
\begin{eqnarray}
x(p)=\frac{L}{\sqrt{p}}  \quad {\rm{and}} \quad y(p)=L\sqrt{p}\;.
\end{eqnarray}

In some cases it is useful to assume there is a starting condition at time $t=0$ defined by
\begin{eqnarray}
x_0 y_0 =L^2
\end{eqnarray}
and price $p_0=y_0/x_0$. This allows to express the token numbers as a function of price according to
\begin{eqnarray}
x(p)=x_0\sqrt{\frac{p_0}{p}}  \quad {\rm{and}} \quad y(p)=x_0\sqrt{p_0 p}\;.
\end{eqnarray}

\subsection{Price dynamics: Brownian Motion and Geometric Brownian Motion}

In stochastic processes, Brownian motion (BM) and Geometric Brownian motion (GBM) serve as fundamental models to describe random behavior in systems such as financial markets (strictly speaking, GBM is more widely used for reasons detailed below but BM offers some advantages in analytical analysis). Both processes are driven by volatility, a parameter that determines the magnitude of random fluctuations, and their statistical behavior is shaped by how this volatility manifests itself. While they share some common properties, they exhibit distinct dynamics due to their underlying mathematical formulations, as detailed below.

\subsubsection{Brownian Motion}

Brownian motion (BM), also known as a Wiener process, is a continuous-time stochastic process characterized by random additive changes at each time step. It models the erratic movement of particles in a fluid or, to some extent, the random fluctuations of asset prices. The evolution of Brownian motion is governed by the stochastic differential equation (SDE)
\[
dP_t = P_0\sigma dW_t,
\]
where $P_t$ is the price at time $t$, $P_0$ is the starting price, $\sigma$ is the volatility, and $W_t$ represents the Wiener process. The parameter $\sigma$ controls the intensity of the random fluctuations: higher volatility implies larger potential deviations in the price over time.

In a discrete setting, BM can be approximated by the simple update rule
\[
P_{t+1} = P_t + P_0\sigma \Delta W,
\]
where $\Delta W$ represents a normal distributed variable. The key feature of this model is that the change in price is additive and driven by volatility, which dictates the magnitude of the randomness at each time step.

\subsubsection{Geometric Brownian Motion}

Geometric Brownian motion (GBM) is a modification of standard Brownian motion, often used to model financial asset prices. GBM follows a multiplicative process, where the price grows or shrinks by a random factor in each time step. Its continuous-time evolution is described by the SDE:
\[
dP_t = \mu P_t dt +P_t \sigma  dW_t,
\]
where $\mu$ represents the drift (mean rate of return) and $\sigma$ is the volatility. Here, volatility still plays a crucial role, but it acts multiplicatively on the current price $P_t$. This means that large prices will experience larger fluctuations than smaller ones, as volatility now scales with the value of the asset.

In the following discussion we neglect $\mu$ without loss of generality. Again, we look for a discrete approximation that is easy to implement in the framework of a simulation.For GBM we use
\[
P_{t+1} = P_t  (1 + \sigma \Delta W),
\]
with volatility $\sigma$ and $\Delta W$ is again a random variable. The use of multiplicative updates ensures that prices remain positive and can exhibit exponential growth or decay over time, depending on the drift and random fluctuations. 

\begin{center}
\begin{figure}
\includegraphics[width=\textwidth]{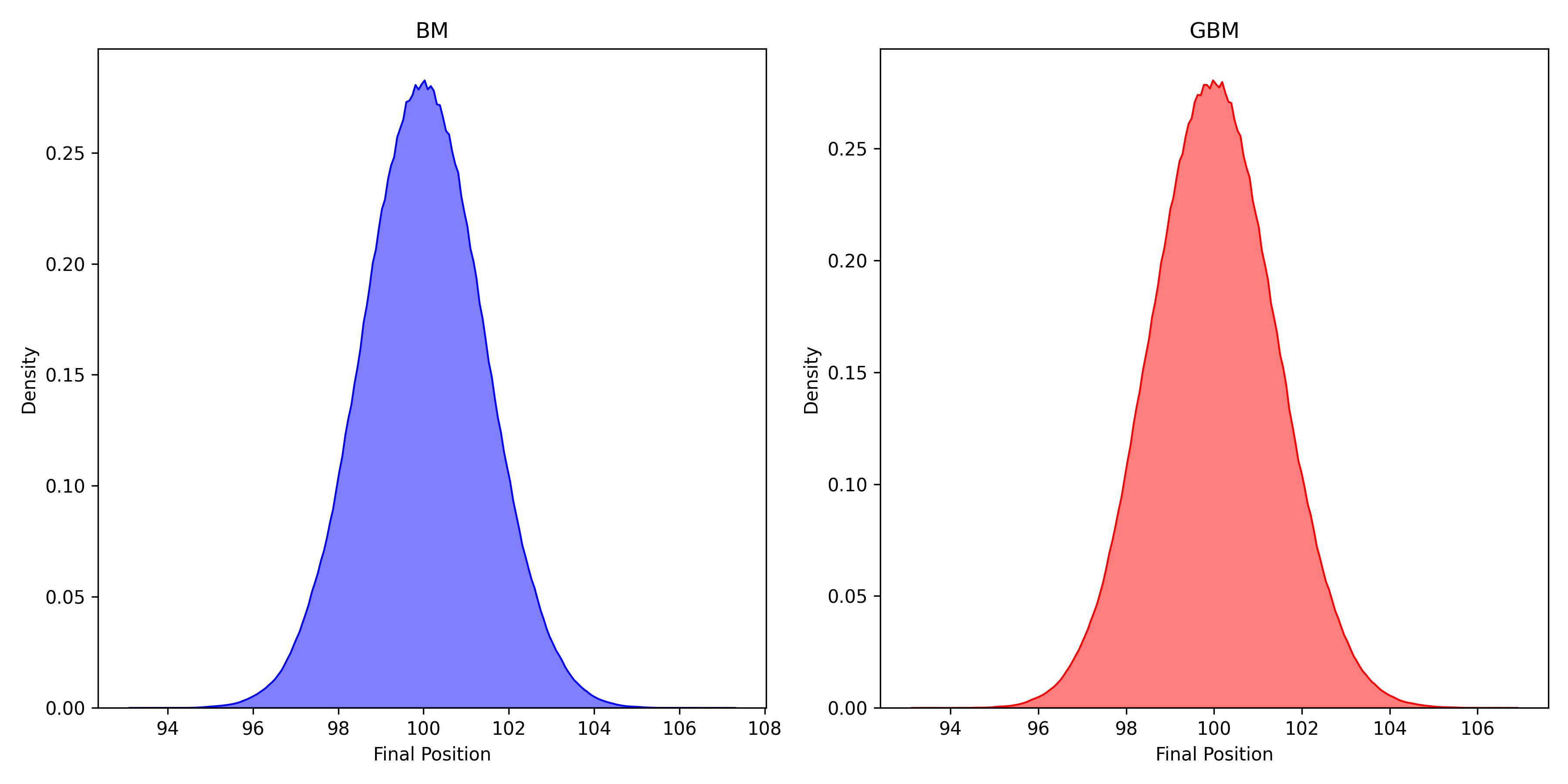}\caption{For short times ($\sigma^2 t$), the distribution functions of final positions from BM and GBM agree very well. The simulations were done for initial price of $P_0=100$, $200$ steps at relative volatility $\sigma=0.001$, and for $40000$ runs.}\label{fig:bmvsgbmshort}
\end{figure}
\end{center}

\begin{center}
\begin{figure}
\includegraphics[width=\textwidth]{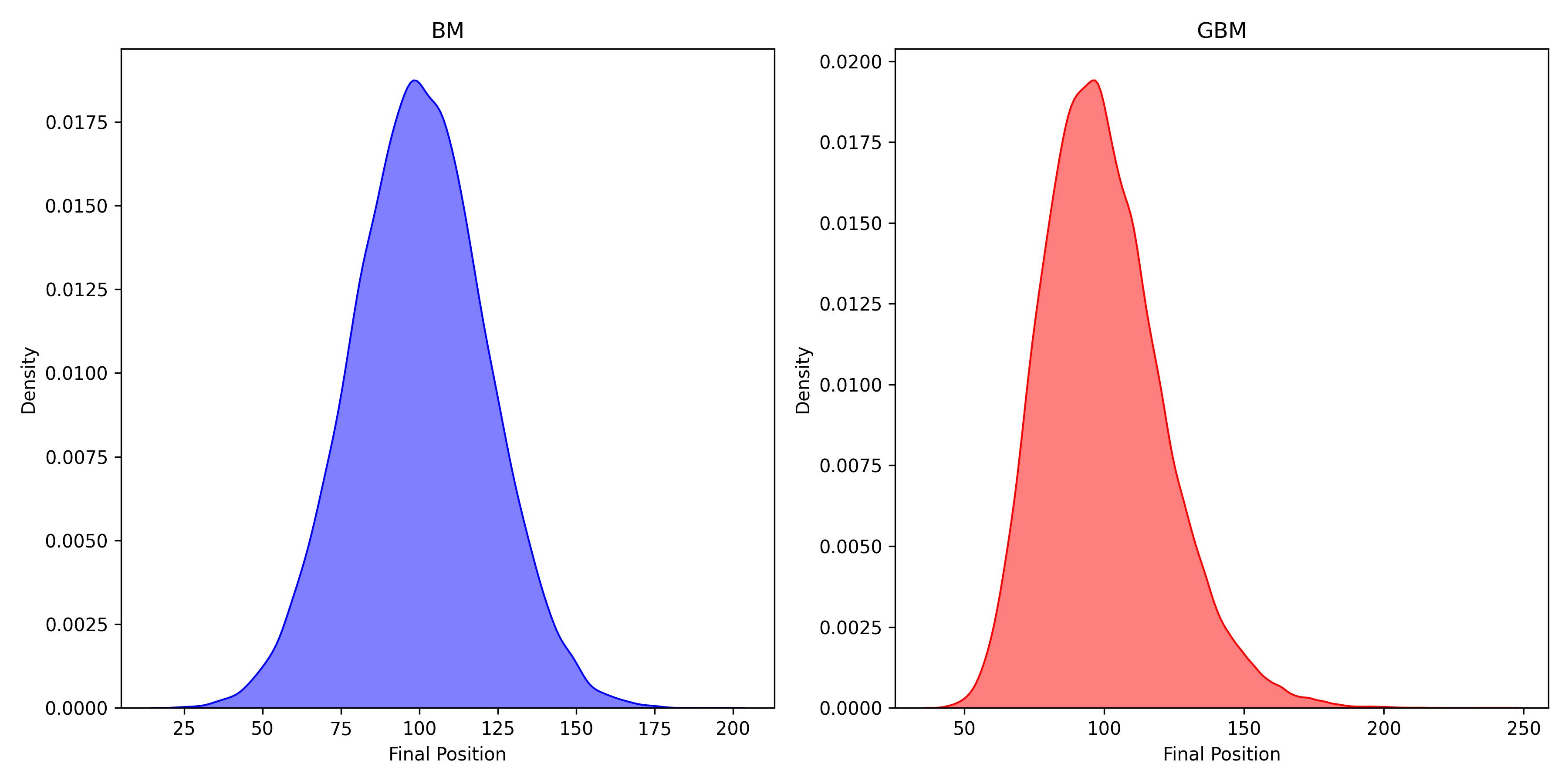}\caption{For longer times ($\\sigma^2t =\mathcal{O}(1)$), the distribution functions of final positions from BM and GBM deviate considerably. The simulations where done for initial price of $P_0=100$, $200$ steps at relative volatility $\sigma=0.015$, and for $40000$ runs.}\label{fig:bmvsgbmlong}
\end{figure}
\end{center}
\subsubsection{Continuum-Time Probability Distributions}

The probability distributions of both BM and GBM in continuous time reflect the role of volatility in shaping their behavior. Volatility ($\sigma$) determines the spread of these distributions and the uncertainty of the future price.

The probability distribution of Brownian motion at time $t$ is a Gauss distribution with mean $P_0$ (the initial price) and variance $\sigma^2 t$. The probability density function (PDF) for Brownian motion is:
\[
\rho_{\rm{BM}}(P_t) = \frac{1}{\sqrt{2\pi P_0^2 \sigma^2 t}} \exp\left(-\frac{(P_t - P_0)^2}{2 P_0^2\sigma^2 t}\right).
\]
Volatility $\sigma$ controls the spread of the Bell curve. A higher volatility increases the standard deviation $\sigma \sqrt{t}$, leading to greater uncertainty in the future price. The key characteristics of this distribution include:
\begin{itemize}
    \item \textbf{Mean}: The mean of the distribution remains constant at $E[P_t] = P_0$.
    \item \textbf{Variance}: The variance grows linearly with time, $\text{Var}(P_t) = P_0^2\sigma^2 t$, showing that the uncertainty increases over time as volatility influences the price spread.
    \item \textbf{Symmetry}: The distribution is symmetric around the initial price, with an equal chance of increasing or decreasing values based on the random increments.
\end{itemize}

In contrast, the price distribution of Geometric Brownian motion is log-normal, a result of the multiplicative nature of the process. The logarithm of the price, $\log(P_t)$, follows a normal distribution with mean $(\mu - \frac{\sigma^2}{2})t + \log(P_0)$ and variance $\sigma^2 t$. The probability density function for GBM is:
\[
\rho_{\rm{GBM}}(P_t) = \frac{1}{P_t \sigma \sqrt{2\pi t}} \exp\left(-\frac{(\log(P_t/P_0) - (\mu - \frac{\sigma^2}{2})t)^2}{2\sigma^2 t}\right).
\]
Here, volatility $\sigma$ again dictates the uncertainty of future prices, but in a multiplicative manner. Larger prices result in proportionally larger deviations due to volatility, which can cause exponential growth or decay.

Key features of the GBM distribution include:
\begin{itemize}
    \item \textbf{Mean}: The mean grows exponentially as $E[P_t] = P_0 e^{\mu t}$, with volatility influencing the rate of spread.
    \item \textbf{Variance}: The variance grows exponentially as well, with larger volatility causing a wider spread of potential outcomes over time.
    \item \textbf{Positivity}: Since GBM ensures positive values, the log-normal distribution reflects this, unlike the symmetric Gaussian distribution of standard Brownian motion.
\end{itemize}

\subsubsection{Common and Distinct Properties}

While both Brownian motion and Geometric Brownian motion are driven by random processes influenced by volatility, there are several key similarities and differences:

\textbf{Common Properties:}
\begin{itemize}
    \item Both processes are driven by random fluctuations determined by volatility $\sigma$, which governs the magnitude of the changes in price over time.
    \item The variance (and hence uncertainty) in both processes increases with time, reflecting greater unpredictability in the future as time progresses.
    \item \textbf{Identical Behavior for Short Time Scales:} Over short time intervals, both Brownian motion and Geometric Brownian motion exhibit similar behavior. For small time steps $\Delta t$, the drift term in GBM becomes negligible, and the dominant term is the stochastic part involving $\sigma dW_t$. This causes both processes to behave identically at short time scales, as the effect of volatility is similar. For a visualization see Fig.~\ref{fig:bmvsgbmshort}.
\end{itemize}

\textbf{Distinct Properties:}
\begin{itemize}
    \item \textbf{Additive vs. Multiplicative Increments:} Brownian motion follows an additive process where price increments are determined by volatility, while GBM follows a multiplicative process, where the relative changes in price depend on both volatility and the current price level.
    \item \textbf{Mean Behavior:} The expected value of a Brownian motion remains constant over time, while the expected value of GBM grows exponentially (unless $\mu=0$), see Fig.~\ref{fig:bmvsgbmlong}.
    \item \textbf{Probability Distributions:} The price distribution in Brownian motion is Gaussian, symmetric around the initial value, while in GBM, the distribution is log-normal, ensuring positive values and allowing for exponential growth, see Fig.~\ref{fig:bmvsgbmlong}.
    \item \textbf{Applicability to Financial Models:} GBM is widely used in financial models for assets, as it reflects the behavior of positive prices and growth due to volatility. In contrast, Brownian motion can result in negative prices.
\end{itemize}

We mostly present this distinction for pedagogical reasons. Henceforth, we will use the GBM update rule for numerical simulations. For analytical calculations as shown in the appendix we use the PDF of BM for convenience and to illustrate how some well-known literature results can be derived in a relatively straightforward manner.

\section{IL vs LVR: Some background}\label{sec:ilvslvr}

A common topic in the context of automated market makers (AMMs) is the relationship between impermanent loss (IL) and loss-versus-rebalancing (LVR). Both are often used as performance metrics for LPs. In this section we introduce the philosophy behind the two metrics and give a simple mathematical formulation in the context of a constant function market maker (the discussion can easily be extended to other types of market makers such as concentrated liquidity). 

\subsection{Impermanent Loss (IL)}

IL measures the difference between the value of a LPers position inside an AMM and the value the LP would have if they had simply held the assets outside the AMM (i.e., a HODL strategy). We assume that the position was made at time $t$ with a corresponding price $p$. The value of the LP's position is given by
\[
V(p) = \frac{L}{\sqrt{p}} + \frac{1}{p}L\sqrt{p} = 2 \frac{L}{\sqrt{p}},
\]
where $L$ represents the initial liquidity provided to the LP (this is characteristic of the position that was initialized).

In the meantime, the price undergoes some dynamics and eventually reaches a price $p_f$ after some time $T$. The value of the position now is
\[
V(p_f) = 2 \frac{L}{\sqrt{p_f}} = 2 \frac{L}{\sqrt{p}} \sqrt{\frac{p}{p_f}}.
\]
In contrast, if the LP had simply held the assets, the value of the position before the price change, called $\rm{HODL}$ from now on, reads:
\[
\text{HODL}(p) = V(p)\;,
\]
while after the price change, the HODL value is
\[
\text{HODL}(p_f) = \frac{L}{\sqrt{p}} \left( 1 + \frac{p}{p_f} \right)\;.
\]
We define IL incurred between $p$ and $p_f$ (or $t$ and $t+T$) as the difference between the hypothetical HODL position and the actual value of the position according to
\begin{eqnarray}\label{eq:IL}
{\rm{IL}}(p, p_f) &=& \text{HODL}(p_f)-V(p_f)\nonumber \\ &=& \frac{L}{\sqrt{p}} \left( 1 - \sqrt{\frac{p}{p_f}} \right)^2 >0\;.
\end{eqnarray}
It is important to note that, irrespective of $p_f>p$ of $p_f<p$, IL is positive. We also see that the whole concept is only sensitive to start and end point of the trajectory and does not care about the exact way to get there. One immediate question after this consideration is: If my position lost value during the price change, who has the corresponding money? This leads to the concept of LVR in a very natural way.

\subsection{Loss-Versus-Rebalancing (LVR)}

Loss-versus-rebalancing (LVR) addresses a question that at first glance sounds different. It examines how the value of a portfolio changes if, instead of being deposited in the AMM, the LPer maintains a shadow portfolio that exactly mirrors the hypothetic LP position over time. This sounds like doing the same thing but there is an important difference. An LP position buys the coins at a price that is inbetween the spot price and the end price, whereas for LVR we assume we can always buy at the end price. What LVR does in addition is that it rebalances after every price change. In that sense, as a quantity, it is sensitive to the exact price trajectory in contrast to IL. We start with considering a differential version of it. We assume that there is a natural time scale $\Delta t$ for a price change, for instance the block time. This implies we track LVR from $t\to t+\Delta t$ and assume there is a corresponding change in price according to $p \to p+\Delta p$

{\bf{After}} the price changes, the LP's position in the AMM has changed according to
\[
x(p + \Delta p) = \frac{L}{\sqrt{p}} \sqrt{\frac{p}{p + \Delta p}}\quad {\rm{and}} \quad y(p + \Delta p) = L \sqrt{p} \sqrt{\frac{p + \Delta p}{p}}\;,
\]
if we concentrate on the individual tokens $x$ and $y$.
To maintain a shadow portfolio that mimics the AMM position but always buys at the end price, the portfolio has to rebalance by buying $\Delta y$ tokens:
\[
\Delta y = y(p + \Delta p) - y(p) = L \sqrt{p} \left( \sqrt{\frac{p + \Delta p}{p}} - 1 \right)\;,
\]
at the price $p + \Delta p$. This requires spending
\[
\Delta \bar{x} = \frac{L}{\sqrt{p}} \left( \sqrt{\frac{p}{p + \Delta p}} - \frac{p}{p + \Delta p} \right)
\]
of the portfolio.
The change in the LP’s token $x$ position in the AMM is
\[
\Delta x = \frac{L}{\sqrt{p}} \left( 1 - \sqrt{\frac{p}{p + \Delta p}} \right)\;,
\]
which turns out to be greater than the cost required to buy the additional $\Delta y$ on the open market, or, $\Delta x> \Delta \bar{x}$. The savings in terms of rebalancing, or the LVR, is therefore
\[
\Delta {\rm{LVR}}(p, p + \Delta p) = \frac{L}{\sqrt{p}} \left( 1 - \sqrt{\frac{p}{p + \Delta p}} \right)^2 >0\;.
\]
One possible interpretation is that the loss of the LPer can be traced back to the fact that the LPer is selling at the wrong price. Any curve based AMM is selling at a better price than the spot price it has after the trade. This immediately leads to two LVR mitigation strategies that are both easier said than done: (1) Decrease liquidity to increase price impact which can soften the impact but also makes the position unattractive for random noise trading. (2) Selling at the end price of the trade, and not at some intermediate price. 

It is important to note that this is the differential version of LVR and the real LVR adds up all the differential versions of it along a price trajectory.

\section{Arbitrage in an AMM without fees: IL, LVR, and arbitrage volume}\label{sec:arbnofees}

Arbitrage with a larger liquidity source is a natural source for trading activity and associated price changes of an AMM. The principle is simple: a trader sees a pricing inefficiency and capitalizes on it. 
We assume that the price of the AMM follows an external infinite liquidity source/oracle at a distance of one time step (block time is a natural candidate for this). Additionally, the arbitrageur does not pay any transaction fees. This is an unrealistic assumption but for a general modeling we take that approach. Furthermore, we assume there is no additional uninformed trading activity. 

We discuss three different periods $T$ over which we track the AMM. The most important distinction is between the first regime and the other two: 

\begin{enumerate}
\item{A short time regime in which $\sigma^2 T\ll 1$.}
\item{An intermediate time regime with $0\ll \sigma^2T<1$.}
\item{A long time limit where $\sigma^2T\geq 1$.}
\end{enumerate}

\subsection{Short time limit}

We first consider the limit of very short times, meaning $T \to 0$. This could for instance be the rice evolution over one single block, or, in other words, {\bf{one step}} of a price change $p \to p + \Delta p$ during $t \to t+T$. In that case we can identify the final price $p_f$ with $p_f=p+\Delta p$ in Eq.~\eqref{eq:IL}. This leads to identical IL and LVR given by
\[
\frac{L}{\sqrt{p}} \left( 1 - \sqrt{\frac{p}{p + \Delta p}} \right)^2\;.
\]
This corroborates that for small changes in a temporal sense IL and LVR are just two different points of view on the same thing, as one would naively expect. 

\subsection{Intermediate time regime}

This is the limit that most of our analysis is concerned with. We consider time evolution during a period $T$, provided that $0 \ll \sigma^2 T <1$ (in this limit, BM and GBM are identical for most practical purposes).
We established above that for infinitesimal times, IL and LVR are identical. Over longer time, they diverge for two reasons: (1) IL is measured between starting point and end point and insensitive to anything that happened in between. (2) LVR adds up IL after every time step along the price trajectory and also resets the reference point through rebalancing after every time step. An extreme example is a price path that returns to its starting point, meaning $p_f=p$ after $T$: IL of such a path is identically zero while LVR is not since it added up and reset after every step. This is in line with the often heard statement: 'impermanent loss only becomes permanent if you rebalance or withdraw'. 

\begin{figure}
\begin{center}
\includegraphics[width=\textwidth]{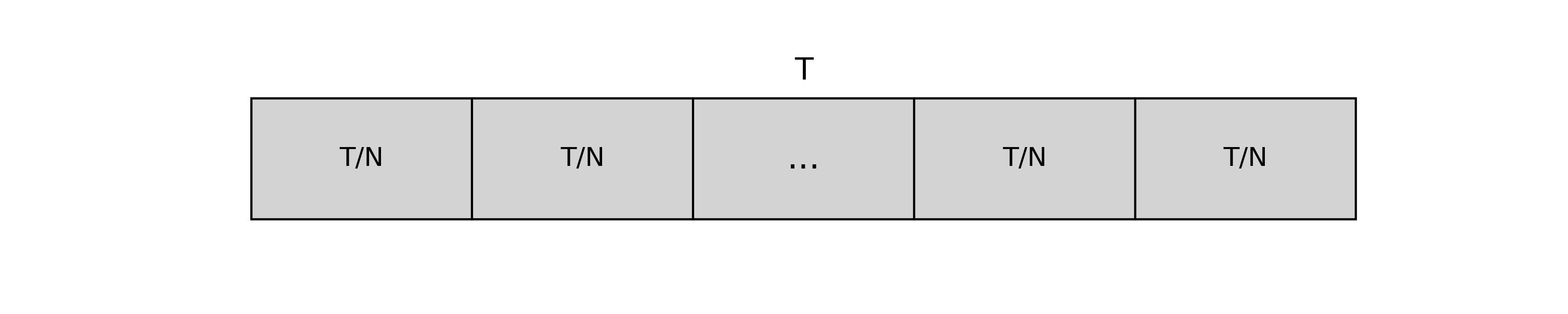}\caption{Throughout the paper, we fix the time frame $T$ over which we track the price dynamics and then chop it into $N$ time steps of equal length $\Delta t=T/N$.}\label{fig:discretization}
\end{center}
\end{figure}

\begin{figure}
\begin{center}
\includegraphics[width=\textwidth]{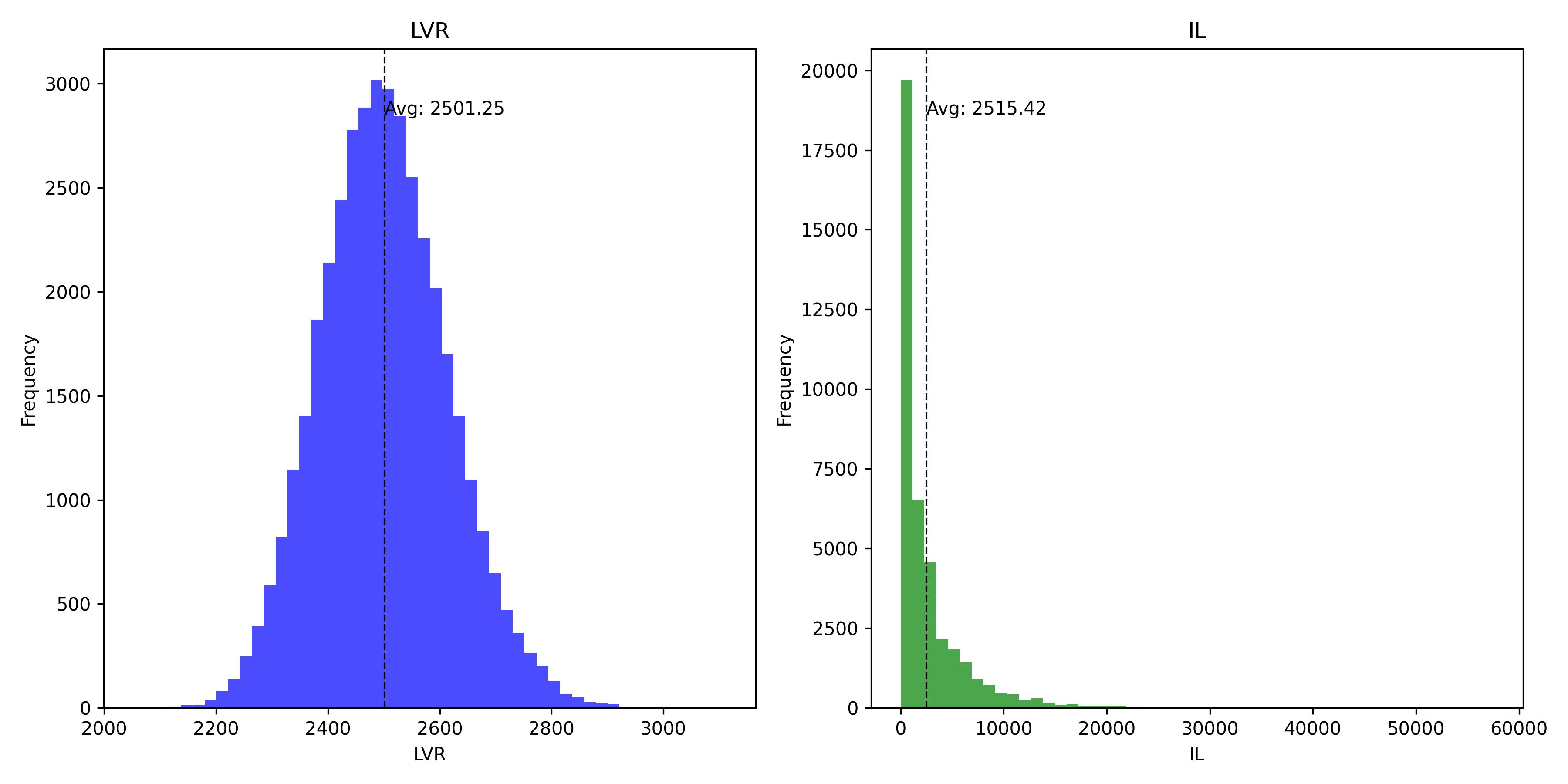}\caption{Histogram of LVR and IL of $40000$ runs performed at initial price $p_{\rm{init}}=100$, liquidity $L=10000$, $\sigma=0.001$, and for $1000$ steps. The averages agree within statistical errors whereas the distribution functions vary significantly.}\label{fig:lvrilnofee}
\end{center}
\end{figure}

Throughout this paper, we consider the time evolution over a total time $T$. We chop this time up into $N$ pieces of equal length, $\Delta t=T/N$. In pratice, this could for instance be block time, see Fig.~\ref{fig:discretization}.
We start our mathematical discussion with some analytical considerations. It is a fair assumption that, for small enough time steps $\Delta t$, the price undergoes a small correction $p \to +\Delta p$ with $\Delta p \ll p$. This allows to expand the expression for LVR during time $\Delta t$ according to
\[
\Delta {\rm{LVR}}(p, p + \Delta p) = \frac{L}{\sqrt{p}} \left( 1 - \sqrt{\frac{p}{p + \Delta p}} \right)^2 \approx \frac{L}{4} \frac{\Delta p^2}{p^{5/2}}\;.
\]
It is important to note that the expectation value of $\langle \Delta p^2 \rangle=p^2 \sigma^2 \Delta t $ for an Ito process. This implies that during $\Delta t$, the average LVR varies according to  
\begin{eqnarray}
\langle \Delta {\rm{LVR}}(\Delta t) \rangle \approx \frac{L}{4\sqrt{p}} \sigma^2 \Delta t= \frac{L}{4\sqrt{p}} \sigma^2 \frac{T}{N}\;.
\end{eqnarray}
This has to be added up over the whole duration of the simulation $T$, meaning over $N$ time steps. On average, this results in 
\begin{eqnarray}\label{eq:lvr}
 \langle {\rm{LVR}}(T) \rangle \approx \frac{L}{4\sqrt{p}} \sigma^2 \frac{T}{N} N=\frac{L}{4\sqrt{p}} \sigma^2 T \;
\end{eqnarray}
if we assume that the time evolution follows a random process as detailed in Sec.~\ref{sec:setup}.
An important observation is that this expression is well behaved in the continuous time limit $\lim T/N \to 0$ (we will encounter the arbitrage volume later as a quantity for which this is not the case). Also, there is another underlying assumption which is that $p$ does not change significantly over the time frame $T$, which is justified if $\sigma^2 T < 1$. Importantly, Eq.~\eqref{eq:lvr} is an expectation value and actual trajectories fluctuate around that value. To illustrate this, we simulated  $40000$ runs with a starting price of $p_{\rm{init}}=100$, liquidity $L=10000$, volatility $\sigma_0=0.001$ for $1000$ steps. The result is shown on the l.h.s of Fig.~\ref{fig:lvrilnofee}. We find that the distribution of LVR is relatively narrow. The intuition is that every step sums up basically the same thing and consequently most trajectories have close to average LVR. This is a direct manifestation of the central limit theorem. We are summing up $N$ positive random numbers, something we explore in more detail below.

IL, on the other hand, is not sensitive to the elementary step within the period $\Delta t$, only to the difference from start to finish, called $\Delta P=p_f-p$. Therefore, we can estimate it using the same formula we used for LVR and find that for a period of length $T$ we get 
\begin{eqnarray}
\langle  {\rm{IL}}(T) \rangle \propto \Delta P^2 =p^2\sigma^2 T \;,
\end{eqnarray}
as an expectation value, just like LVR, see Appendix. If we perform the same simulation we performed for LVR, we find a very different scenario. While LVR and IL have identical average values, their distribution functions differ significantly. Most trajectories have very small IL. This simply reflects the fact that the distribution of final prices of all price trajectories is centered at the starting price in the absence of drift. 
One can determine the distribution function of IL analytically in a straightforward manner. One has to express the price in terms of IL, meaning one has to identify $p+\Delta p$ with $p(t)$ and invert Eq.~\eqref{eq:IL} bringing it into the form $p({\rm{IL}})$. The full distribution function can then be obtained from 
\begin{eqnarray}
\left|\frac{dp_t ({\rm{IL}})}{d {\rm{IL}}}\right|\rho_{\rm{BM/GBM}}(p_t({\rm{IL}}))=\bar{\rho}_{\rm{BM/GBM}}({\rm{IL}})
\end{eqnarray}
where $\bar{\rho}_{\rm{BM/GBM}}({\rm{IL}})$ is the distribution function of IL after time $t$ and for a given volatility $\sigma$ (to be more precise, one has to worry about integral boundaries and double solutions but this describes the general strategy). The factor $|dP_t ({\rm{IL}})/d {\rm{IL}}|$ corresponds to the Jacobian associated with the integral measure and results from the transformation of variables. The full expression is a bit bulky and given by
\begin{eqnarray}\label{eq:ildist}
\bar{\rho}_{\rm{BM/GBM}}({\rm{IL}})&=&\frac{1}{\sqrt{{\rm{IL}}}}\frac{p_0^{5/4}}{\sqrt{L}} \frac{\rho_{\rm{BM/GBM}}\left(p_0/\left(p_0^{1/4}\sqrt{{\rm{IL}}/L}+1 \right)^2\right)}{\left(1+p_0^{1/4}\sqrt{{\rm{IL}}/L}\right)^3} \nonumber \\ &+&\frac{1}{\sqrt{{\rm{IL}}}}\frac{p_0^{5/4}}{\sqrt{L}} \frac{\rho_{\rm{BM/GBM}}\left(p_0/\left(1-p_0^{1/4}\sqrt{{\rm{IL}}/L} \right)^2\right)}{\left(1-p_0^{1/4}\sqrt{{\rm{IL}}/L}\right)^3}\theta \left( L/\sqrt{p_0}-{\rm{IL}}\right)\;.\nonumber \\
\end{eqnarray}
More details including an explicit derivation are provided in the Appendix where we show how to carry out the steps.

In the limit of small IL we find an approximate form for the distribution function (one could as well expand the full solution, but we prefer a more intuitive derivation here). For now we introduce $p'=p-p_0$, which is a small deviation from the average price. For low IL, we have 
\begin{eqnarray}
{\rm{IL}} \propto p'^2\;.
\end{eqnarray}
Consequently, we can express the price in terms of IL according to
\begin{eqnarray}
p({\rm{IL}})=\pm \sqrt{{\rm{IL}}}\;.
\end{eqnarray}
Using this expression to derive the Jacobian, we find that the distribution function for small IL reads
\begin{eqnarray}
\bar{\rho}_{\rm{BM}}({\rm{IL}})\approx \frac{a}{\sqrt{{\rm{IL}}}}e^{-c {\rm{IL}}/(2 \sigma^2 t)}
\end{eqnarray}
with $a,c$ being constants composed of $L$, $p_0$, $\sigma$, and $t$. The constants can be easily obtained from expanding the exact distribution function shown in the Appendix to the order shown above. For GBM, we find a similar expression, again not shown here.

However, the main message is that the distribution function has a divergence $1/\sqrt{{\rm{IL}}}$ in the limit of small IL, in agreement with our numerical findings in Fig.~\ref{fig:lvrilnofee}.
This implies that most trajectories perform better than the average value when it comes to IL. On the other hand, there are statistical outliers that generate huge IL. One can also determine the expectation value of IL after time $T$ according to
\begin{eqnarray}
\langle {\rm{IL}} \rangle = \int_{0}^{\infty}d {\rm{IL}} {\rm{IL}} \bar{\rho}_{\rm{BM}}({\rm{IL}})=\frac{L}{4\sqrt{p_0}}\sigma^2 T\;,
\end{eqnarray}
in agreement with the result shown in the Appendix. 

\subsubsection{LVR distribution function as central limit theorem result of the IL distribution}

\begin{figure}[h]
    \centering
    \includegraphics[width=0.9\textwidth]{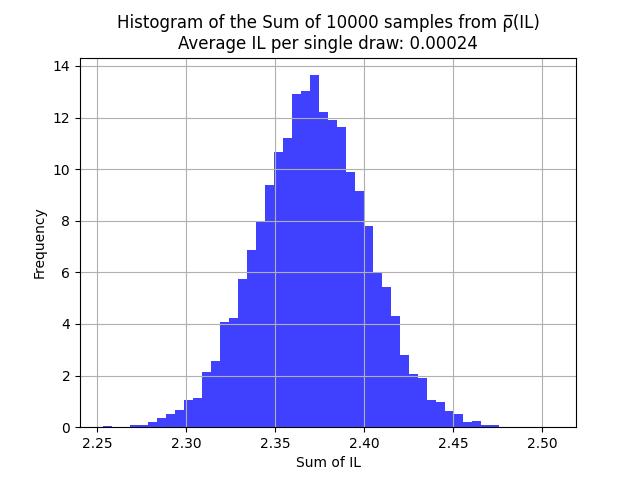}
    \caption{ \textbf{Histogram of the Sum of 10000 Samples from the Distribution $\bar{\rho}(\mathrm{IL})$} This histogram represents the distribution of the sum of 5000 samples drawn from the custom probability distribution $\bar{\rho}(\mathrm{IL})$, where the IL values are sampled according to the normalized weights from the function. The distribution is derived from a Gaussian-modified form with specific parameters $p_0 = 100$, $L = 10000$, $\sigma = 0.1$, and $t = 1.0$. The frequency of sums is plotted, with 50 bins representing the density of summed values. We find that the average IL per draw agrees with the result presented in the previous subsection as well as in the appendix for the given values.    }
    \label{fig:sumil}
\end{figure}

We argued above that for small time steps, IL and LVR are the same. In that sense, LVR must be equivalent to the process of adding up many instances of infinitesimal IL randomly drawn according to the distribution function, Eq.~\eqref{eq:ildist}. Strictly speaking, one has to rebalance the reference point at each step. However, in the intermediate time regime this can be neglected to an excellent degree of approximation. To show this explicitly, we draw $10000$ random values of IL according to its distribution function (we chose $L=10000$, $p_0=100$, $t=1$, and $\sigma=0.1$ for better visibility) and repeat the experiment $1000$ times. The is shown in the histogram in Fig.~\ref{fig:sumil}. It shows a Gaussian distribution in agreement with the central limit theorem. The total LVR agrees with the value predicted in the previous section and the Appendix.

\subsection{Long time limit}

\begin{figure}
\begin{center}
\includegraphics[width=\textwidth]{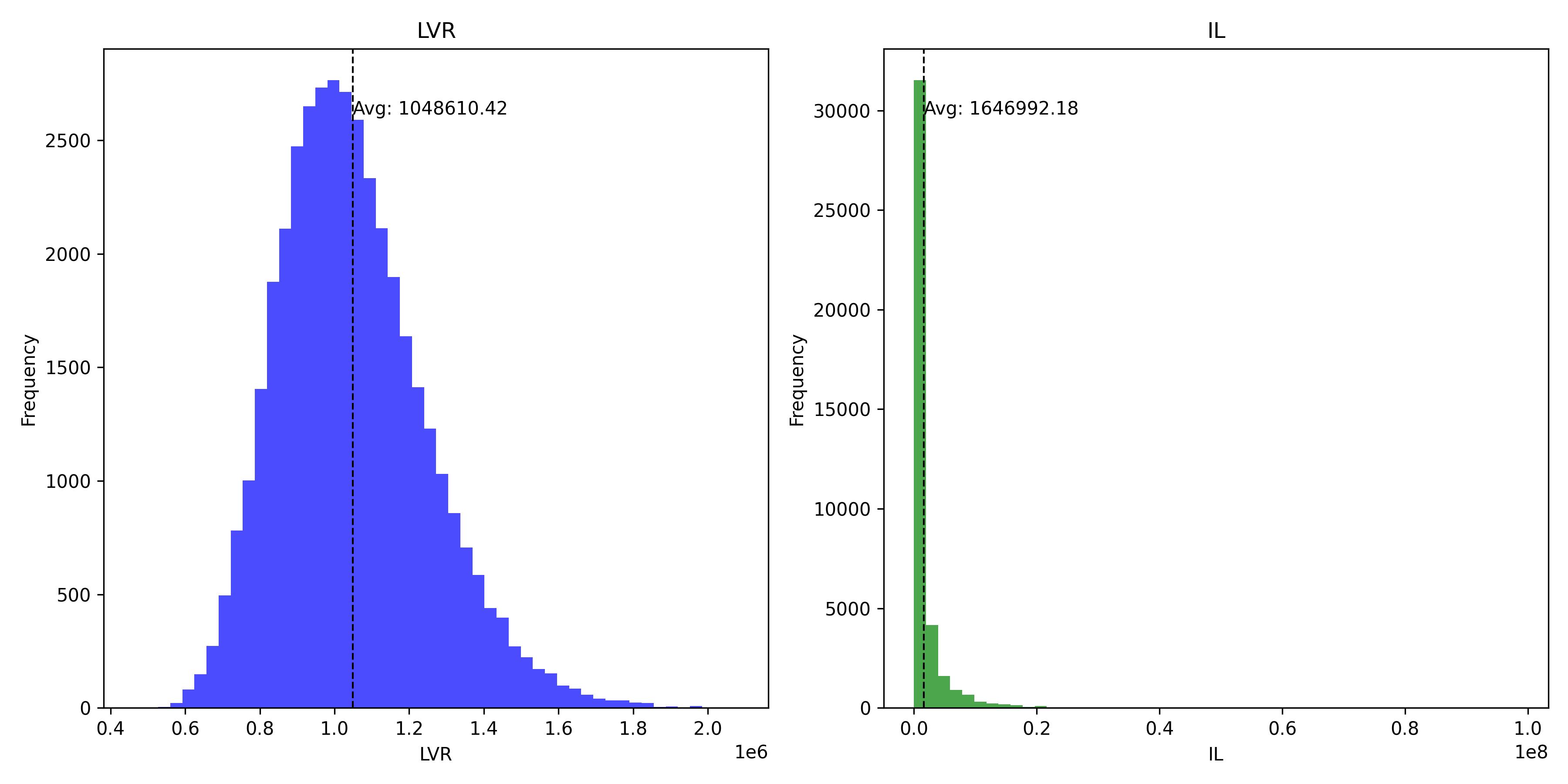}\caption{Histogram of LVR and IL of $10000$ runs performed at initial price $p_{\rm{init}}=100$, liquidity $L=10000$, $\sigma=0.02$, and for $1000$ steps. The averages of the distribution functions differ significantly.}\label{fig:lvrlongtime}
\end{center}
\end{figure}

For short and intermediate times, the distinction between Brownian and geometric Brownian motion turns out to be irrelevant, and the previous sections considered that limit. However, for longer times, or, equivalently, larger volatility, the differences become pronounced. To be more precise, we are looking at the limit $\sigma^2 T>1$. To showcase this, Fig.~\ref{fig:lvrlongtime} shows a simulation that is done for geometric Brownian motion with $p_{\rm{init}}=100$, $L=10000$, $\sigma=0.02$, $1000$ recorded over $10000$ runs. We observe that the averages of the LVR and IL distributions differ significantly in that situation. Furthermore, the distribution function of LVR looks skewed like a log-normal distribution. We have checked the literature to which extent it can be expected that LVR, which is the sum of many instances of infinitesimal IL that follows a probability distribution shown in Fig.~\ref{fig:lvrlongtime} r.h.s. produce a true log-normal distribution. The belief in the literature is that adding random numbers following a truncated power-law distribution, like the one observed for IL, can produce approximate log-normal distributions over ranges, but not precisely. 

\subsection{Summary of LVR vs IL}

In order to discuss the relationship between IL and LVR it is important to distinguish three time regimes:

\begin{enumerate}
\item{A short time regime in which $\sigma^2 T\ll 1$.}
\item{An intermediate time regime with $0\ll \sigma^2T<1$.}
\item{A long time limit where $\sigma^2T\geq 1$.}
\end{enumerate}

In the very short time regime, there is no difference between LVR and IL. In the intermediate-time regime, there is a marked difference in the statistical properties of the distribution functions. In that regime, it is important to note that the price evolution is not yet sensitive to the difference between BM and GBM. We find that LVR has a normal shaped distribution while IL has a distribution that becomes power-law singular for small IL. However, both distributions have the same average value (shown analytically in the Appendix), and the distribution function of LVR can be constructed from the sum of many values of infinitesimal IL drawn from its distribution function in accordance with the central limit theorem. For longer times, the GBM becomes significant, and the distribution functions differ more strongly. The LVR distribution function itself starts to resemble that of a log-normal distribution, whereas the IL distribution function does not change significantly. However, the average values also start to deviate significantly. All these findings are comprehensively summarized in Tab.~\ref{tab:illvr}

\begin{table}[h!]
\centering
\begin{tabular}{|c|c|c|}
\hline
\textbf{Short Time} & \textbf{Intermediate Time} & \textbf{Long Time} \\ \hline
$\sigma^2 T \ll 1$ & $0 \ll \sigma^2 T < 1$ & $\sigma^2 T \geq 1$ \\ \hline
$\rm{IL} = \rm{LVR}$ & $\rm{IL} \neq \rm{LVR}$ & $\rm{IL} \neq \rm{LVR}$ \\ \hline
 & $\langle \rm{IL} \rangle = \langle \rm{LVR} \rangle$ & $\langle \rm{IL} \rangle \neq \langle \rm{LVR} \rangle$ \\ \hline
\end{tabular}
\caption{Relationship between IL and LVR over different time regimes.}\label{tab:illvr}
\end{table}
The remainder of the paper is concerned with the time regime termed 'intermediate'.

\subsection{Arbitrage volume}

\begin{figure}
\begin{center}
\includegraphics[width=0.8\textwidth]{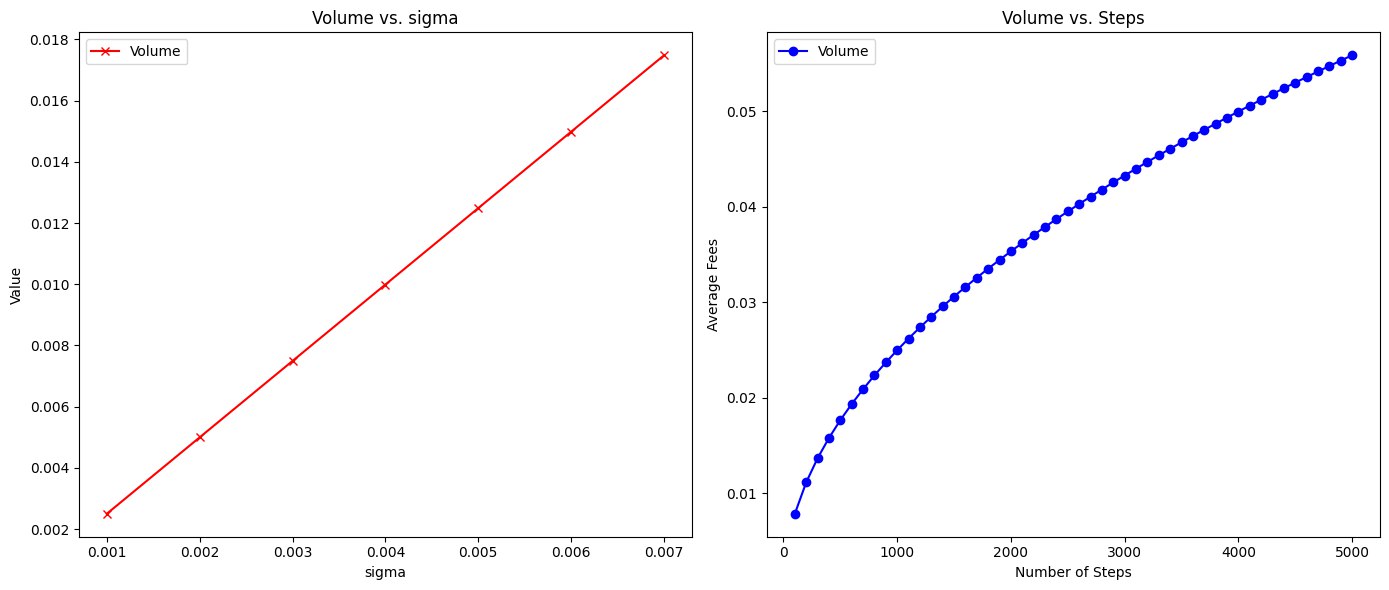}\caption{The volume shows linear behavior with increasing $\sigma$ (l.h.s) as well as a square root divergence upon increasing the discretization steps (l.h.s), both in agreement with Eq.~\eqref{eq:vol}}\label{fig:volsim}
\end{center}
\end{figure}

Another interesting quantity to look at is the volume that results from arbitraging the price difference from $p \to p+\Delta p$. The corresponding flow of tokens is given by
\begin{eqnarray}
\Delta {\rm{Vol}}(p,p+\Delta p)=\left|x(p+\Delta p)-x(p)\right|=\frac{L}{\sqrt{p}}\left|1-\sqrt{\frac{p}{p+\Delta p}}\right| \approx \frac{L}{2p^{3/2}}|\Delta p|\;.
\end{eqnarray}
On average, this implies
\begin{eqnarray}
\langle \Delta {\rm{Vol}}(p,p+\Delta p)\rangle \approx \frac{L}{2p^{3/2}}\langle |\Delta p| \rangle\;.
\end{eqnarray}
For BM/GBM, this leads to
\begin{eqnarray}
\langle \Delta {\rm{Vol}} \rangle \propto p\sigma \sqrt{\Delta t}\;,
\end{eqnarray}
meaning it is linear instead of quadratic in the volatility. This has an important consequence: For $\Delta t \to 0$, the average arbitrage volume diverges and its continuum limit is not well defined. 
Let's investigate this in more detail. Again, we consider the evolution over a total time $T$. Now, we chop this time up into $N$ pieces of equal length, $\Delta t=T/N$. This implies that during that time the volume varies as 
\begin{eqnarray}
\langle \Delta {\rm{Vol}}(\Delta t) \rangle \propto p \sigma \sqrt{\frac{T}{N}}
\end{eqnarray}
which has to be summed up over the whole duration, meaning N pieces. This leads to 
\begin{eqnarray}\label{eq:vol}
\langle {\rm{Vol}}(T) \rangle \propto p \sigma \sqrt{\frac{T}{N}}N=p \sigma \sqrt{T} \sqrt{N}\;.
\end{eqnarray}
In order to confirm this, we have run a series of simulations. We performed two types: (I) We fixed the number of steps in the simulation and track the volume as a function of $\sigma$. This results in a linear behavior as shown in Fig.~\ref{fig:volsim} l.h.s, in agreement with the analytical estimate, Eq.~\eqref{eq:vol}; (II) We fix the total volatility over a time-frame $T$ and vary the number of steps $N$ into which we split up $T$. We find that at fixed $\sigma^2 T$, the volume diverges as $\sqrt{N}\sigma \sqrt{T}$, see Fig.~\ref{fig:volsim} r.h.s., again in perfect agreement with the prediction. We finish this discussion with three comments: (1) The limit $\lim T/N \to 0$ is artificial and in reality there will always be a finite block time or time associated with transactions that sets a lower bound for $\Delta t$. (2) In practice, there are also fees and transaction costs that serve as a cutoff for short-time behavior, as we will discuss in the following section. (3) Even in the absence of fees, there is the tick spacing of the underlying pool which provides a natural cutoff for this divergence. 

\section{Arbitrage with fees}\label{sec:illvrfee}
The previous section established the dynamics of IL and LVR in the absence of fees, highlighting their sensitivity to price fluctuations and volatility. We now turn our attention to the presence of fees. This alters arbitrage dynamics fundamentally by creating a no-trade region. This adjustment impacts both LVR and IL, albeit with differing degrees of effectiveness, as explored below.

The following discussion, apart from the distribution functions (to the best of our knowledge they have not been discussed in the literature to this point), is mostly a more pedestrian reformulation of some of the results of Ref.~\cite{milionis2023automated}.

\begin{figure}
\centering
\includegraphics[width=0.9\textwidth]{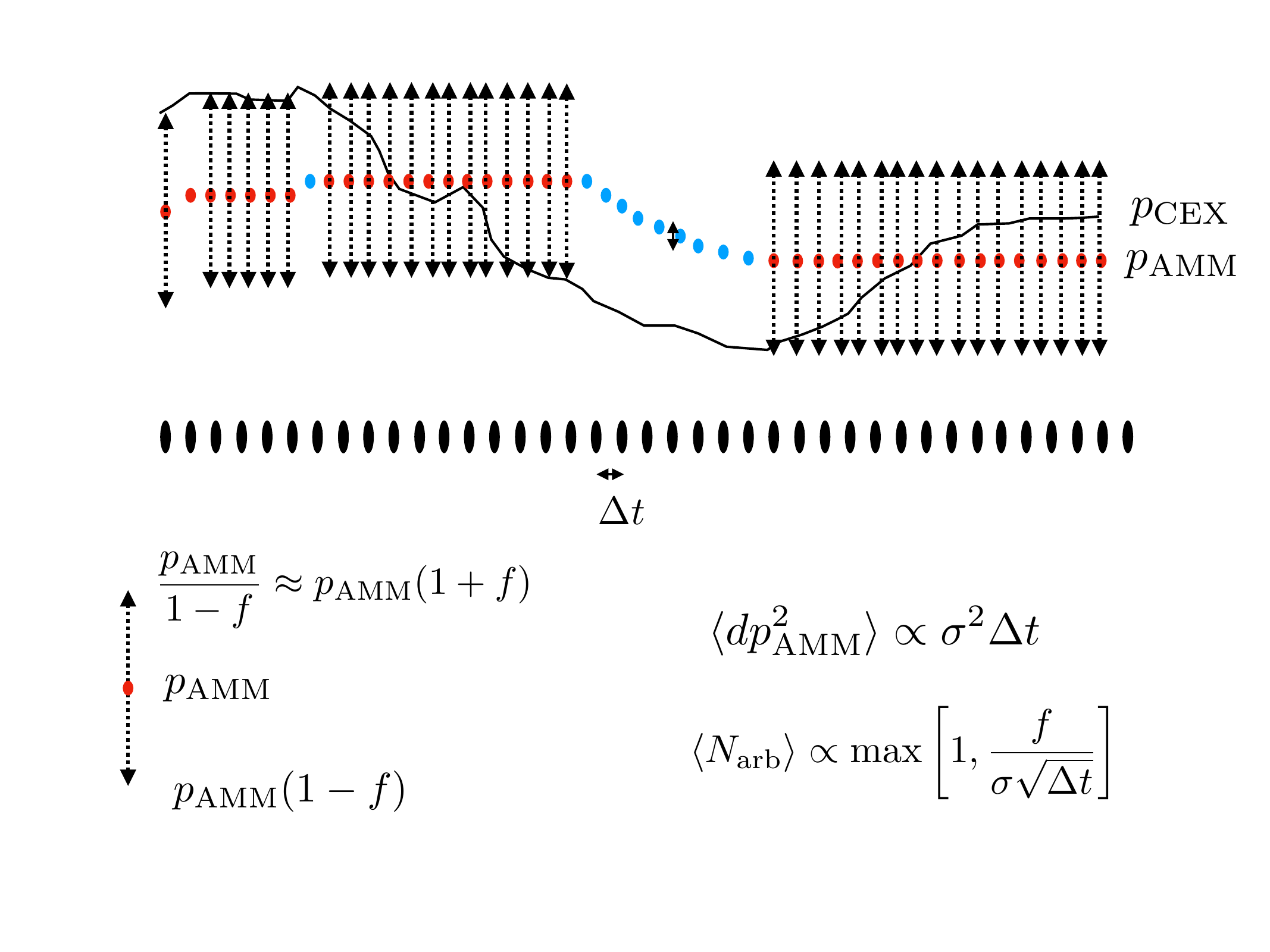}\caption{Upper panel: Fees create a now-arbitrage zone around the current price of the AMM. The black line denotes the price of a CEX whereas the red dots denote the price of the AMM with the black dotted arrows showing the non-arbitrage zone. Arbitrage is 'checked' on a time scale $\Delta t$. After every successful arbitrage event, there can either be an immediate next atomic arbitrage if the price keeps its directional movement, or the price can reverse which delays the subsequent event. Lower panel: There are two different regimes concerning the average number of steps $\langle N_{\rm{arb}}\rangle$, depending on whether the movement during $\Delta t$ is larger than the no-arbitrage region or not. }\label{fig:arbzone}
\end{figure}

\begin{figure}
\centering
\includegraphics[width=0.9\textwidth]{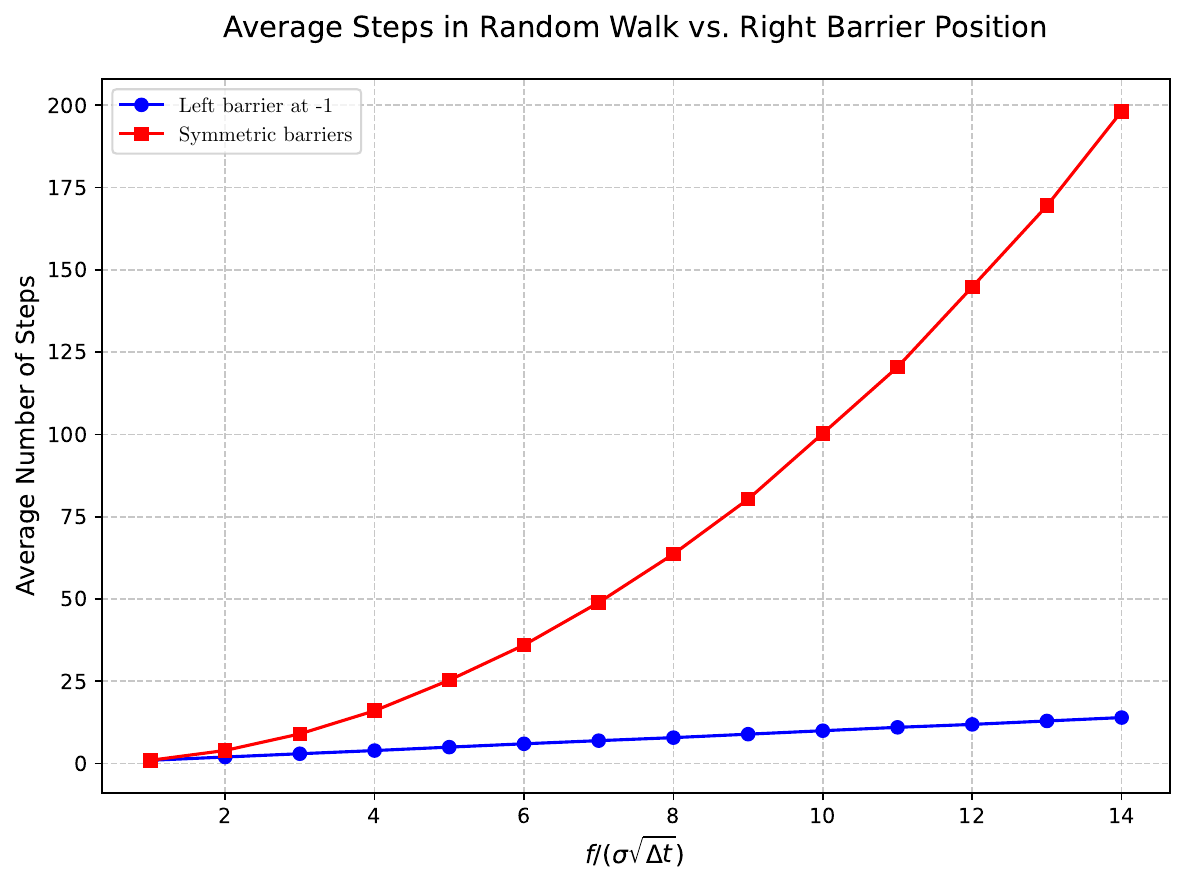}\caption{Average number of steps required to reach a symmetric barrier (red) at $\Delta p_{\rm{rel}}=\pm f/(\sigma \sqrt{\Delta t})$ or one with one barrier at zero and the other at $\Delta p_{\rm{rel}}=-f/(\sigma \sqrt{\Delta t})$ (blue). While in the case of symmetric barriers the number of steps required grows quadratically, in agreement with the expectation for a diffusion process, it grows linearly in the case of asymmetric barriers. }\label{fig:rwbarrier}
\end{figure}

\begin{figure}
\centering
\includegraphics[width=0.9\textwidth]{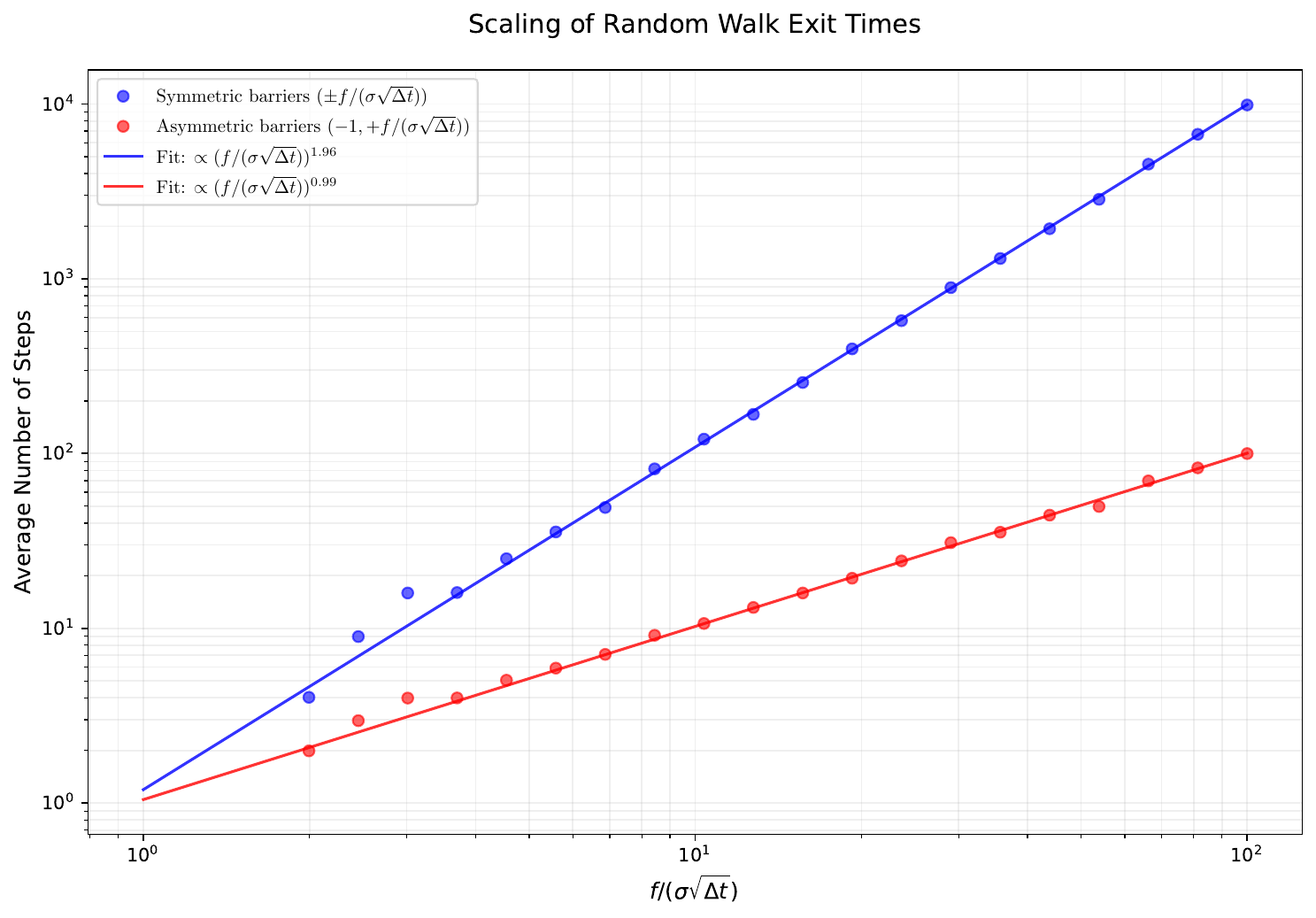}\caption{
Average number of steps required for a random walk to exit under two different barrier configurations: symmetric barriers at $ \pm f / (\sigma \sqrt{\Delta t})$ (blue circles) and asymmetric barriers at $ -1 $ and $ +f / (\sigma \sqrt{\Delta t})$ (red circles). The data is fit to power laws, with the symmetric case scaling as $ \propto (f / (\sigma \sqrt{\Delta t}))^{b}$ (blue line) and the asymmetric case scaling as $ \propto (f / (\sigma \sqrt{\Delta t}))^{b}$ (red line). The fitted exponents show distinct scaling behaviors for the two setups. Error bars have been omitted for clarity.
}\label{fig:rwbarrierlog}
\end{figure}

\begin{figure}
\centering
\includegraphics[width=0.9\textwidth]{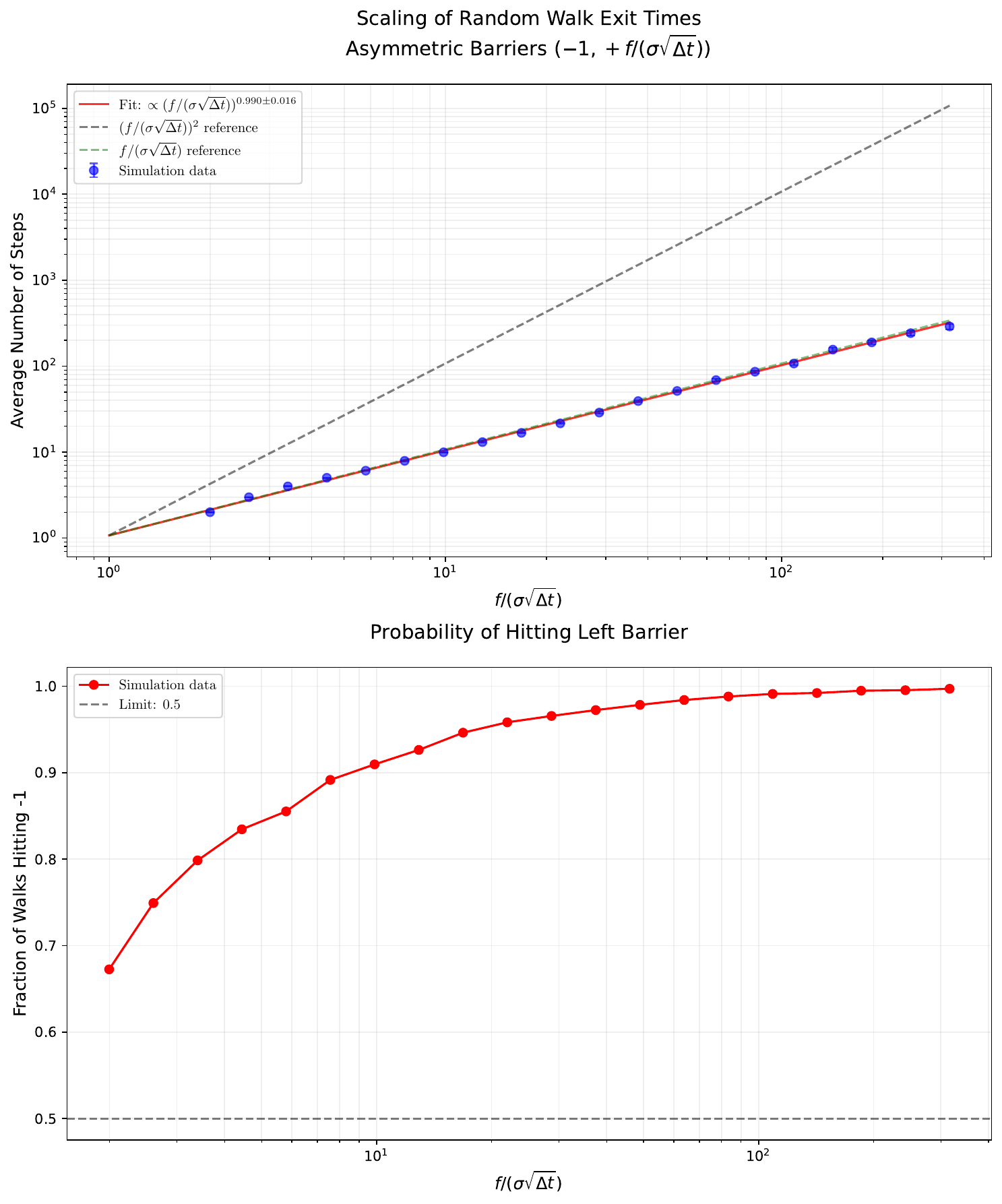}\caption{
(Top) Average number of steps required for a random walk to exit with asymmetric barriers at $-1$ and $+f/(\sigma\sqrt{\Delta t})$, plotted against $f/(\sigma\sqrt{\Delta t})$. The blue circles represent simulation data with error bars showing the standard error. The red line indicates the power law fit, with an exponent of $\tau = (f/(\sigma\sqrt{\Delta t}))^b$, where the fitted value of $b$ is provided with uncertainty. Reference lines for $(f/(\sigma\sqrt{\Delta t}))^2$ and $f/(\sigma\sqrt{\Delta t})$ scaling are shown for comparison. (Bottom) The fraction of random walks that hit the left barrier at $-1$ as a function of $f/(\sigma\sqrt{\Delta t})$. The red curve shows the simulation data, while the dashed black line represents the theoretical limit of 0.5 for large $f$.
}\label{fig:rwbarriermisc}
\end{figure}

The primary role of fees in the dynamics of arbitrage is that they create a region of no arbitrage, (almost) symmetrically around the current price of the AMM $p_{\rm{AMM}}$ with an upper limit $p_{\rm{upper}}=p_{\rm{AMM}}/(1-f)\approx p_{\rm{AMM}}(1+f)$ and the lower limit $p_{\rm{lower}}= p_{\rm{AMM}}(1-f)$. It is important to understand the nature of an arbitrage event. Let us assume there was a successful arbitrage event at time $t$ with a CEX price of $p_{\rm{CEX}}=p_{\rm{AMM}}(1+f)$ (from now on we suppresse the subscript AMM). After a successful arbitrage, we ask: When will the next arbitage event take place? The underlying time structure is such that the next chance to arbitrage is at $t+\Delta t$. The corresponding change in price during that time is $|dp|=p \sigma \sqrt{\Delta t}$ and the corresponding
\begin{eqnarray}
\Delta {\rm{LVR}} &\propto& \sigma^2 \Delta t \nonumber \\ \Delta {\rm{Vol}} &\propto& \sigma \sqrt{\Delta t}\;.
\end{eqnarray}
If the price, $p_{\rm{CEX}}$ goes up during $\Delta t$, there is a guaranteed atomic arbitrage event. 
On the other hand, if the CEX price moves down, there is no immediate arbitrage event. For a falling price one now has to hit $p(1-f)$ for the next arbitrage or return to the point of last arbitrage. This situation is illustrated in Fig.~\ref{fig:arbzone}. We are now going to answer how many steps it will take on average to have an arbitrage event after just having undergone one.  We can answer this question by constructing a random walk that mimic this situation. To warm up, let's first ask the following questions: In a random walk starting at $p$ and with an elementary step size of $\Delta p=p \sigma \sqrt{\Delta t}$, how many steps $\langle N \rangle$ does it take on average to reach a boundary located at $p\pm p \Delta p_{\rm{rel}}$? The answer to this is $\langle N \rangle \propto \Delta p_{\rm{rel}}^2/(\sigma^2 \Delta t)$. This is a direct consequence of the behavior of the mean squared displacement in a diffusion process.
However, to mimic the actual arbitrage situation, we have to modify the question at hand. The correct question is to ask: When, starting from $p$, the price either goes above $p$ or reaches $p-2pf$? This corresponds to a very asymmetric situation. The somewhat surprising answer to this is that the expected number of steps to the next arbitrage event follows
\begin{eqnarray}
\langle N_{\rm{arb}} \rangle \propto \frac{f}{\sigma \sqrt{\Delta t}}\;.
\end{eqnarray}
This can easily be simulated by means of a Monte Carlo simulation that tracks the average number of steps required, shown in Fig.~\ref{fig:rwbarrier}, Fig.~\ref{fig:rwbarrierlog}, and Fig.~\ref{fig:rwbarriermisc}. We explain the details of the simulations in the figure captions. 

To summarize, the characteristic number of steps is
\begin{eqnarray}
\langle N_{\rm{arb}}\rangle=\left\{\begin{array} {cc} 1 & \Delta t \gg \frac{f}{\sigma} \\ \frac{f}{\sigma \sqrt{\Delta t}} &  \Delta t \ll \frac{ f}{\sigma} \end{array}\right.
\end{eqnarray}
The full LVR and Volume are given by 
\begin{eqnarray}
{\rm{LVR}}\propto \sigma^2 \Delta t \frac{N}{\langle N_{\rm{arb}}\rangle}  \propto \left \{\begin{array}{cc} \sigma^2 T & \Delta t \gg \frac{f}{\sigma} \\ \frac{\sigma^3 T}{f} \sqrt{\Delta t} &  \Delta t \ll \frac{f}{\sigma} \end{array} \right.
\end{eqnarray}
\begin{eqnarray} {\rm{Vol}}\propto \sigma \sqrt{\Delta t} \frac{N}{\langle N_{\rm{arb}}\rangle}\propto \left \{\begin{array}{cc} \sigma \sqrt{T} \sqrt{N} & \Delta t \gg \frac{ f}{\sigma} \\ \frac{\sigma^2 T}{pf}  &  \Delta t \ll \frac{ f}{\sigma} \end{array} \right.
\end{eqnarray}

\subsection{Simulations with fees}

\begin{figure}
\begin{center}
\includegraphics[width=\textwidth]{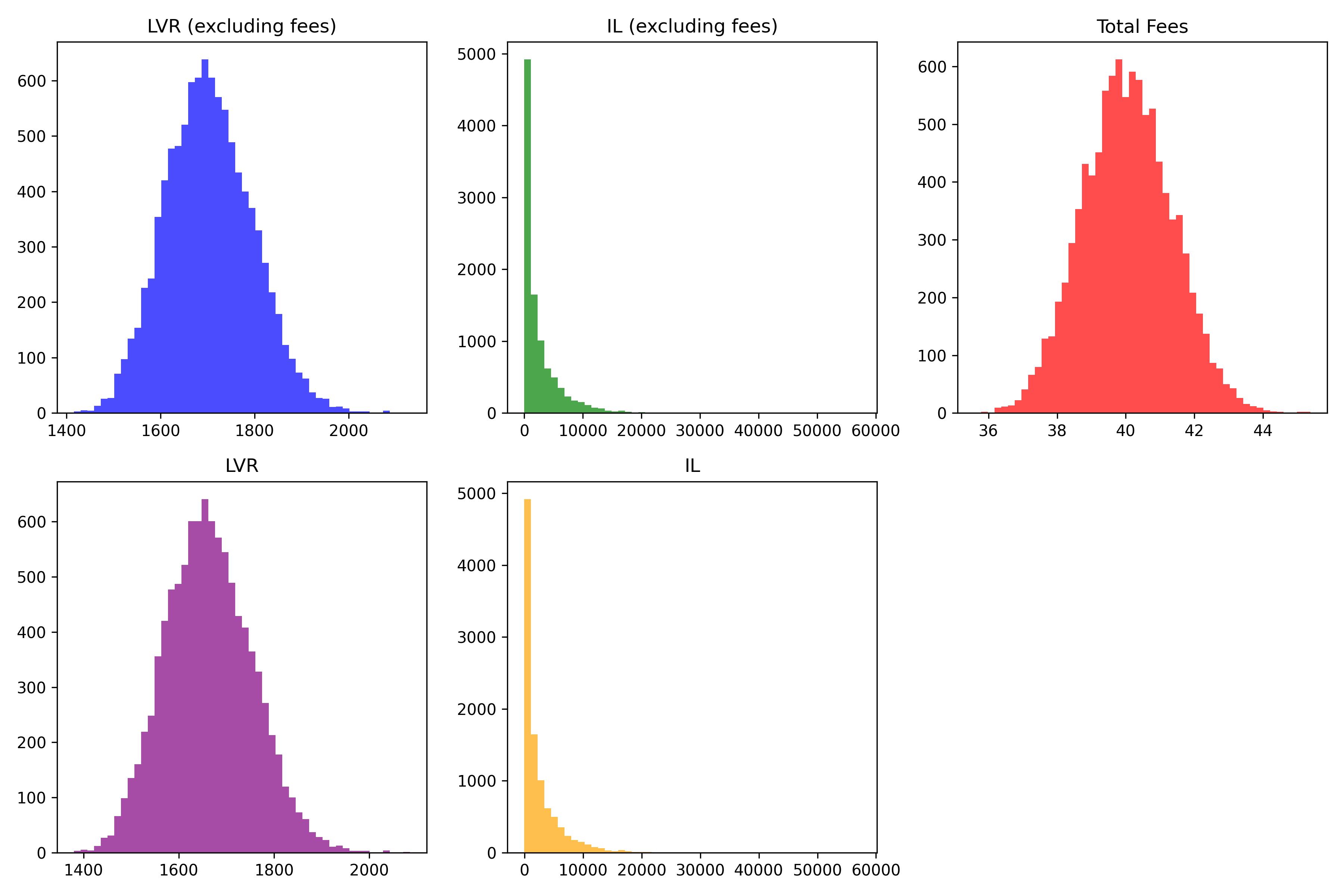}    \caption{Distributions of key metrics from $10000$ simulations of an Automated Market Maker (AMM) over $1000$ time steps. The AMM is initialized with $1000$ tokens of asset A and an equivalent value of asset B at an initial price of $100$. The price evolution follows a discrete geometric Brownian motion with volatility $\sigma = 0.001$. The base fee is set to $0.02\%$. The subplots show histograms of: (a) Loss vs. Rebalancing (LVR) excluding fees, (b) Impermanent Loss (IL) excluding fees, (c) Total fees collected, (d) LVR including fees, and (e) IL including fees. This visualization allows for a comparison of the distributions of LVR and IL both with and without the impact of fees, as well as the distribution of fees collected by the AMM. The differences in these distributions highlight the distinct behaviors of LVR and IL under various market conditions and the impact of fees on liquidity provision in AMMs.}
\label{fig:lvrfee}
\end{center}
\end{figure}

We are now going to discuss how those time scales play out in a full simulation. We have made a simulation of the key metrics, shown in Figure.~\ref{fig:lvrfee}. The AMM is initialized with $1000$ tokens of asset A and an equivalent value of asset B at an initial price of $100$. The price evolution follows a discrete geometric Brownian motion with volatility $\sigma = 0.001$. The base fee is set to $0.02\%$. The subplots show histograms of: (a) Loss vs. Rebalancing (LVR) excluding fees, (b) Impermanent Loss (IL) excluding fees, (c) Total fees collected, (d) LVR including fees, and (e) IL including fees. This visualization allows for a comparison of the distributions of LVR and IL both with and without the impact of fees, as well as the distribution of fees collected by the AMM. The differences in these distributions highlight the distinct behaviors of LVR and IL under various market conditions and the impact of fees on liquidity provision in AMMs. 

We make a number of important observations. The first is that LVR in absence of fees accounted still follows a relatively narrow distribution around an average value. However, the average LVR, is largely reduced compared to the case in the absence of fees (more on that later). IL, on the other hand, looks largely unaltered compared to the case with fees. This statement remains true for the average. So we can first conclude that fees reduce the average LVR substantially more than average IL while they do relatively little to the distribution functions, as expected. The intuition is simple: fees prevent trades from happening thereby reducing LVR. IL, on the other hand, only cares about start and end point of the price trajectory which is largely unaffected. Fees due to arbitrage themselves also have a normal shaped distribution. If they are added to counter LVR we find that LVR is again reduced but not fully mitigated. The intuition is again clear: as long as arbitrage happens, LVR is positive. The only way to counter LVR completely is to prohibit arbitrage, altogether. For IL, the consequences are more positive. The bulk of price trajectories was close to having no IL by virtue of most price trajectories showing average price. This implies that for many price trajectories, the introduction of fees leads to a net positive markout once IL and fees are considered together, even if only adverse trading is considered.  

\begin{figure}
\begin{center}
\includegraphics[width=\textwidth]{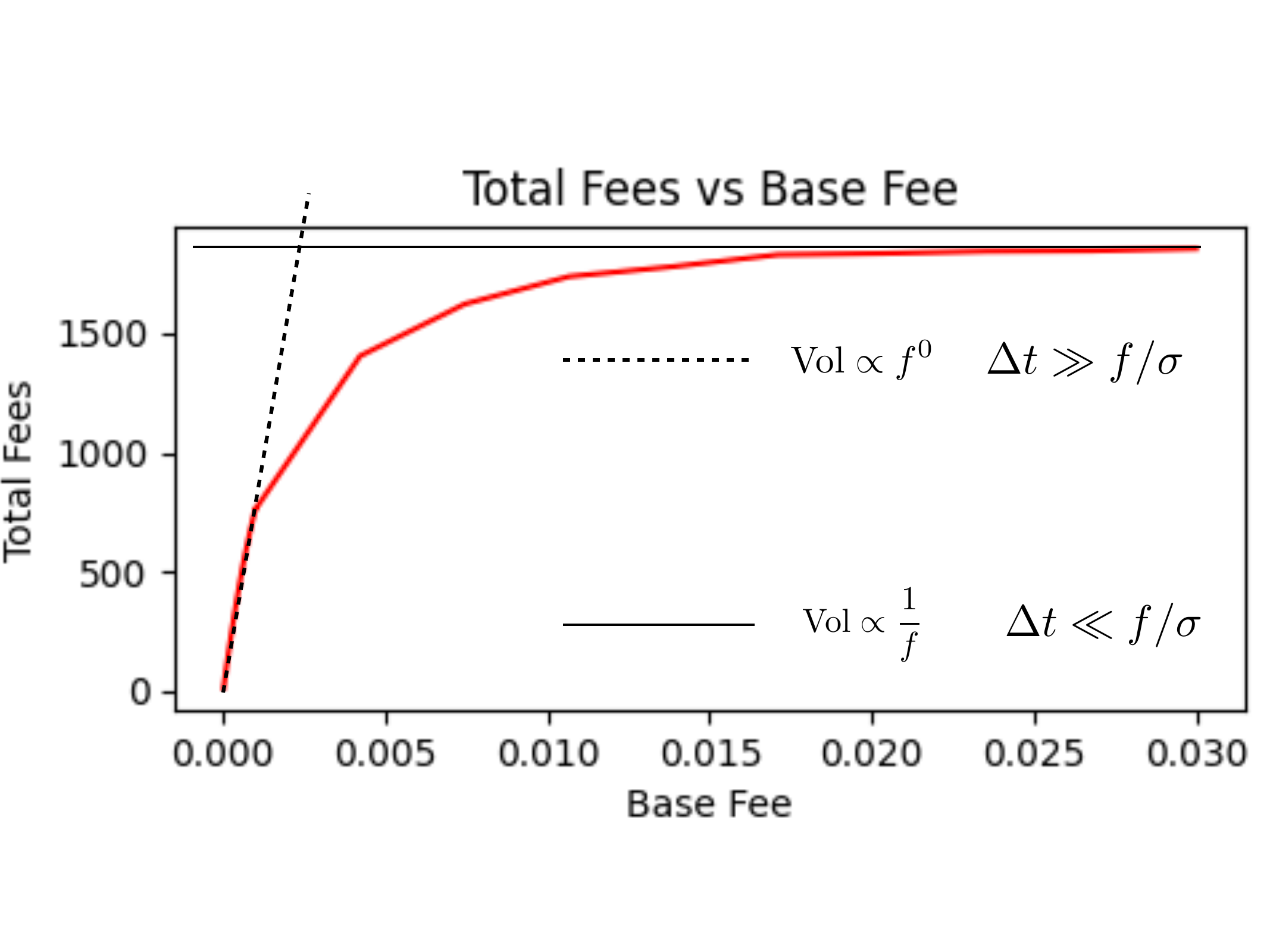}    \caption{Results from $5000$ simulations of an Automated Market Maker (AMM) over $1000$ time steps for a range of base fees. The AMM is initialized with $1000$ tokens of asset A and an equivalent value of asset B at an initial price of $100$. The price follows a discrete geometric Brownian motion with volatility $\sigma = 0.004$.}
\label{fig:volvsfee}
\end{center}
\end{figure}

Let's look into the dynamics of fees some more. In the limit where $\Delta t_{\rm{arb}}\ll\Delta t$, the fees will be $\propto f$ whereas in the limit $\Delta t_{\rm{arb}}\gg\Delta t$ we will find fees $\propto f^0$. The behavior for values for $10000$ simulations of $1000$ steps at a price $p_{\rm{init}}=100$, $\sigma=0.001$, $f=0.0002$, and liquidity $L=10000$ is shown in Fig.~\ref{fig:lvrfee}. A more detailed look at the behavior of volume in the two regions is shown in Fig.~\ref{fig:volvsfee} which shows the different regimes as well as the large crossover region.

\begin{figure}
\begin{center}
\includegraphics[width=\textwidth]{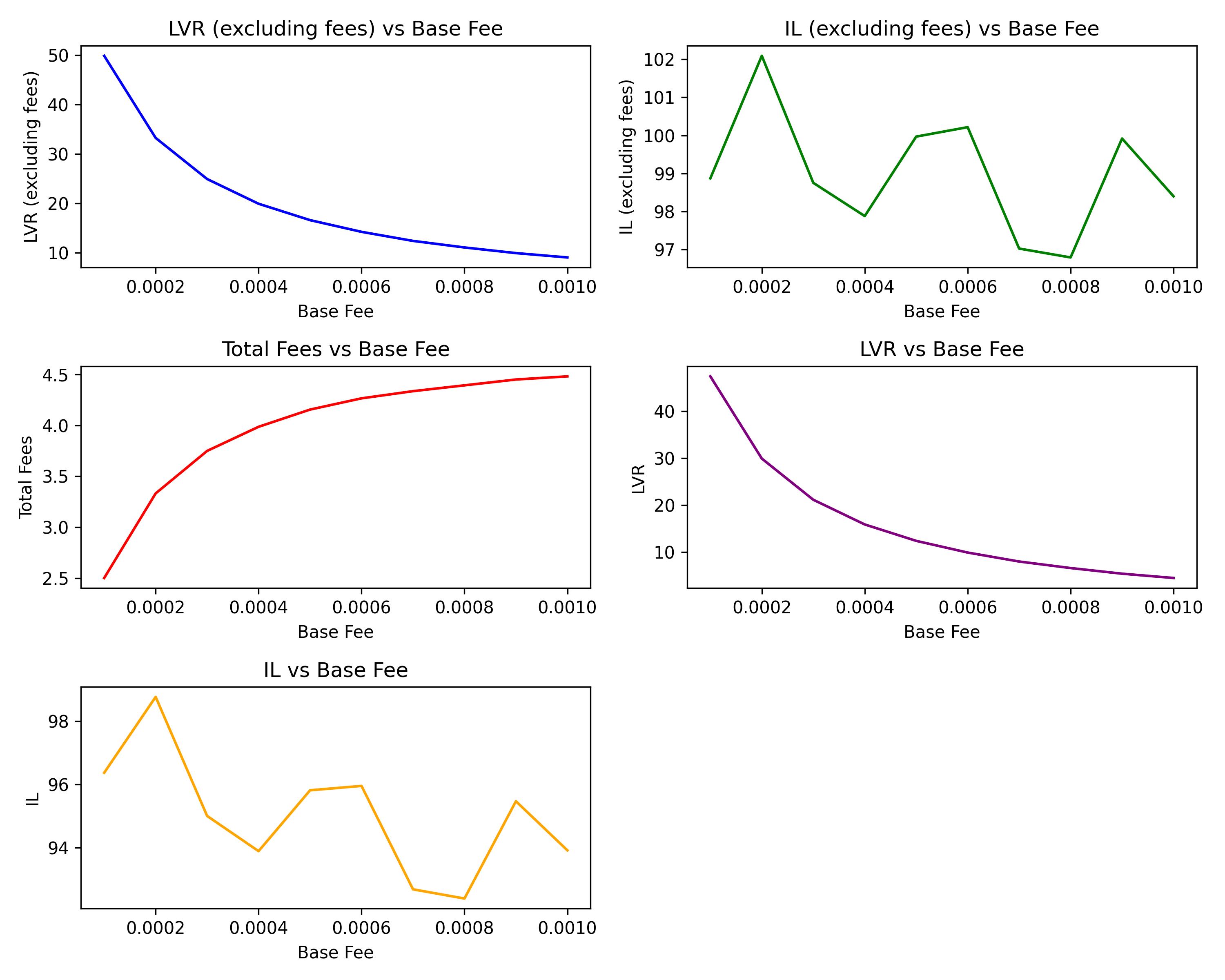}    \caption{Results from $10000$ simulations of an Automated Market Maker (AMM) over $1000$ time steps for a range of base fees. The AMM is initialized with $1000$ tokens of asset A and an equivalent value of asset B at an initial price of $100$. The price follows a discrete geometric Brownian motion with volatility $\sigma = 0.0002$. The subplots show the relationship between the base fee and the following key metrics: (a) Loss vs. Rebalancing (LVR) excluding fees, (b) Impermanent Loss (IL) excluding fees, (c) Total fees collected, (d) LVR including fees, and (e) IL including fees. These plots illustrate how different base fee levels affect LVR, IL, and the total fees collected by the AMM.}
\label{fig:lvrvsfee}
\end{center}
\end{figure}

To finish the discussion we studied the averages of the quantities shown in Figure.~\ref{fig:lvrfee} as a function of the base fee, see Fig.~\ref{fig:lvrvsfee}. As we argued, LVR is very sensitive to the base fee whereas IL shows to be relatively unaffected, as argued above. 

\subsection{Summary}

With regards to an introduction of fees, IL and LVR react very differently: 

LVR gets affected in two ways: (i) the trading activity is reduced by virtue of the arbitrage frequency going down so it coarse grains some of the moves and does not participate. (ii) It benefits from gaining fees, meaning fees help reduce LVR in two ways. 

IL, on the other hand, only benefits in the sense of the second point, meaning by receiving fees. This can best seen in the numerical simulations where we find large reductions of LVR but almost no effect on IL, both on averages as well as distribution functions. However, on the positive side most trajectories have close to zero IL and fees can hel render the markout net positive.

To summarize, fees are a great tool to reduce the impact of informed flow on LVR, but do very little to mitigate the average IL. For that one needs different strategies, such as hedging by means of options. On the other hand, most price trajectories show well below average IL so in many scenarios they are efficient in helping mitigate the effects of adverse trading.

\section{Conclusion}\label{sec:conc}
This paper has explored the intricate relationship between impermanent loss (IL), loss-versus-rebalancing (LVR), and the role of fees in automated market makers (AMMs). Our analysis reveals several key findings:
\begin{enumerate}
\item For infinitesimal price changes, IL and LVR are mathematically identical, despite their different interpretations. However, over longer periods they differ significantly.

\item In the absence of fees, both LVR and IL have the same expectation value in an intermediate time regime. However, they have hugely different distribution functions. While most price trajectories have close-to-average LVR, most price trajectories have well below average IL. 

\item In the absence of fees, both IL and LVR scale linearly with time and quadratically with volatility in the intermediate time regime. The arbitrage volume, on the other hand, scales linearly with volatility for finite time steps and exhibits a square root dependence on the number of time steps, leading to a divergence in the continuous time limit.

\item The introduction of fees creates a no-trade region for arbitrage, effectively introducing a characteristic time scale. This time scale determines whether the system behaves more like the fee-less case (when the arbitrage time is much shorter than the block time) or enters a fee-dominated regime.

\item The presence of fees reduces average LVR much more strongly than average IL by skipping trades. 

\item The fee structure significantly impacts the behavior of arbitrage volume and collected fees, with a transition from fee-proportional to fee-independent regimes as the characteristic arbitrage time exceeds the block time.

\item Fees help mitigate LVR much more efficiently than IL. Mitigating IL needs more refined strategies, such as options, but mitigating LVR definitely helps soften the possible impact.

\item Interestingly, however, by virtue of the extremely skewed distribution function of IL, in most situations LPers will be subject to well below average IL. This implies that fees can turn the net IL after fees are added positive, something that is never possible for LVR.

\end{enumerate}
These findings have important implications for the design and optimization of AMM protocols. They highlight the delicate balance between providing liquidity, managing IL and LVR, and setting appropriate fee structures to ensure the long-term sustainability and efficiency of decentralized exchanges.
A key takeaway from this is that while fees help mitigate LVR, they cannot fully compensate it. However, since IL has a large probability of being below LVR in general, there is a good chance that fees overcompensate IL which is important for closing the position. Consequently, reducing LVR is a good protection against toxic flow losses in general, whether you worry about LVR or IL. They can help reduce LVR substantially and can turn IL even net positive in many situations. Having said that, fees cannot protect against huge IL (which, unfortunately, usually comes with close-to-average LVR) which can only be achieved through additional hedging techniques.

Future research could focus on developing more sophisticated optimal dynamical fee structures that balance liquidity provision incentives with arbitrage deterrence, and investigating the impact of more complex price processes on AMM performance. Additionally, it is imperative to study non-toxic trading in addition to arbitrage as well as including fees and possibly ordering into the analysis. By deepening our understanding of these fundamental mechanisms, we can contribute to the development of more robust and efficient decentralized finance ecosystems, ultimately fostering greater financial inclusion and innovation in the blockchain space.



\section*{Acknowedgment}

We acknowledge useful discussions with A. Elsts, R. Fritsch, Gigasafu, and D. Robinson.

{\small
  \bibliographystyle{ACM-Reference-Format}
  \bibliography{references}
}
\newpage

\appendix

\section{Differential equation for LVR}\label{app:lvrde}

While IL only cares about the start and end points of the price path, LVR is summed up along the whole path. To better describes this, we now convert the expression for LVR into a differential equation. We start with assuming that the changes in price $dp \ll p$ and expand the expression for LVR according to 
\[
\Delta {\rm{LVR}}(p, p + dp) = \frac{L}{\sqrt{p}} \left( 1 - \sqrt{\frac{p}{p + dp}} \right)^2 \approx \frac{L}{4} \frac{dp^2}{p^{5/2}}
\]

For Brownian motion it is well known that $d p^2=\sigma_0 d t$ meaning we 
find
\begin{eqnarray}
\Delta{\rm{LVR}}(p,\Delta t)=\frac{L}{4} \frac{\sigma_0^2 dt}{p^{5/2}}.
\end{eqnarray}
Performing the limit $d t \to 0$ we can convert this into a differential equation according to
\begin{eqnarray}
\frac{d {\rm{LVR}(p(t))}}{dt}=L\frac{\sigma_0^2}{4 p(t)^{5/2}}\;.
\end{eqnarray}
We note that in this differential equation the time dependence is implicit in the trajectory $p(t)$ and therefore depends on the individual realization.

To summarize: IL is calculated between start and end points of an observed time frame and as such does not care about intermediate losses. LVR, on the other hand is updated after every price change on the trajectory. They are connected in the following sense: LVR sums up IL of price changes within the individual time unit $d t$. A naive expectation is that LVR should be much bigger than IL because one is summing up pieces all the time. We will find that this expectation is wrong on average but true for most paths.

We will devote the following section to finding a better understanding of their relation. 

\section{Analysis of IL and LVR}\label{app:illvr}

We use two tools in this section: numerical simulations based on the random walk as well as statistical properties of the Gaussian distribution. We find, as expected, excellent agreement between the two.

\subsection{Random walk analysis}

\begin{figure}[h]
    \centering
    \includegraphics[width=0.8\textwidth]{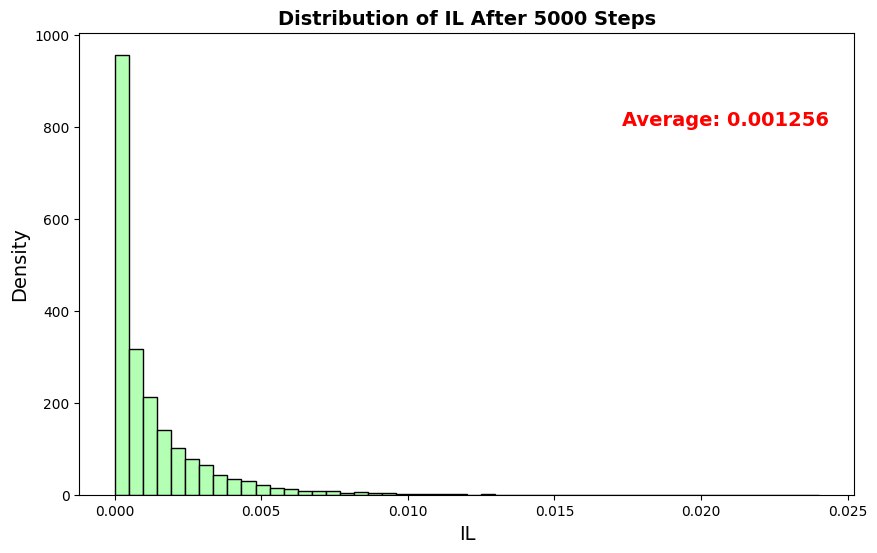}
    \caption{Distribution of IL over 20000 runs.}
    \label{fig:MCILvsLVRa}
\end{figure}

\begin{figure}[h]
    \centering
    \includegraphics[width=0.8\textwidth]{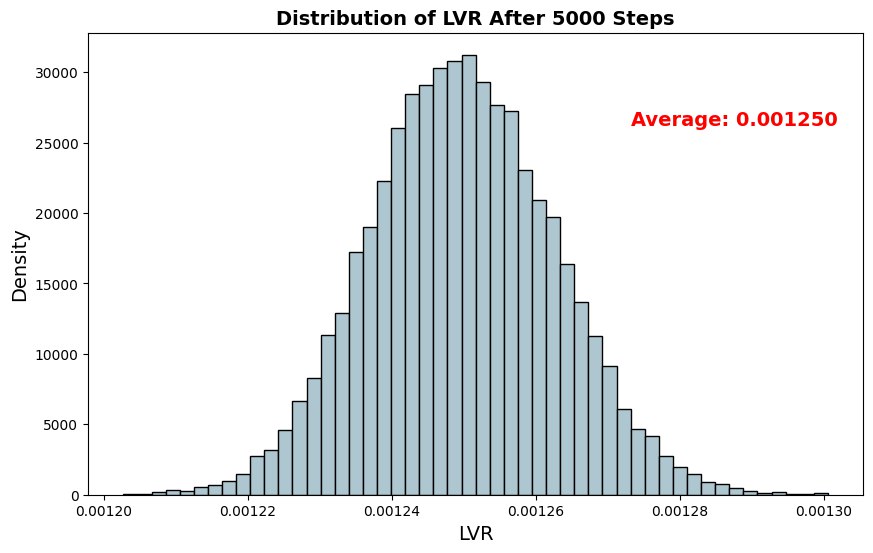}
    \caption{Distribution of IL over 20000 runs.}
    \label{fig:MCILvsLVRb}
\end{figure}
In all the subsequent plots we have chosen the following setting. We choose a starting price of $p_0=100$ and $x_0=100$, as well as $\sigma_0=0.01$. A single run consists of an evolution of $5000$ time steps and we perform 20000 runs. We record both histograms of the runs as well as averages. 

We calculated IL (Fig.~\ref{fig:MCILvsLVRa}) and LVR (Fig.~\ref{fig:MCILvsLVRb}) for the same settings. We find, maybe surprisingly at this point, that while both quantities have very different distribution functions, they possess the same average. This has been verified for many other parameter settings so this is not a coincidence but independent of parameters. 

The shape of the distribution functions can easily be rationalized. IL only measures a loss at the end point relative to the starting point. Most trajectories for the random walk end in a position relatively close to $p_0$. Those trajectories contribute very little IL meaning the majority of trajectories has sub-average IL. LVR, on the other hand, realizes a loss at every step. Since the trajectories predominantly hover around $p_0$, those losses are roughly the same for every trajectory at every time step. Consequently, most trajectories collect average LVR. The surprising insight is that the average IL and the average LVR agree within statistical accuracy. An immediate question is whether one is a more useful metric to quantify losses than the other. The advantage of LVR is that looking at a number of positions gives a good chance to identify the correct value while with IL the bulk of the contributions from trajectories with little probability so IL will easily underestimate the actual loss (if more positions with different starting points were considered). 

We will now use the properties of the Gaussian distribution to show the agreement between the averages is no coincidence.

\subsection{Analytical treatment}

At this point we cannot refrain from stating that the following procedure has a very prominent counterpart in the theory of quantum mechanics, which is the Feynman path integral. The Feynamn path integral sums up all the possible paths that a particle could take to go from one place in space-time to another. If we replace space with price, we have the correspondence (since this ia a classical problem it is in fact more similar to the Wiener integral).

\begin{figure}
\centering
\includegraphics[width=0.8\textwidth]{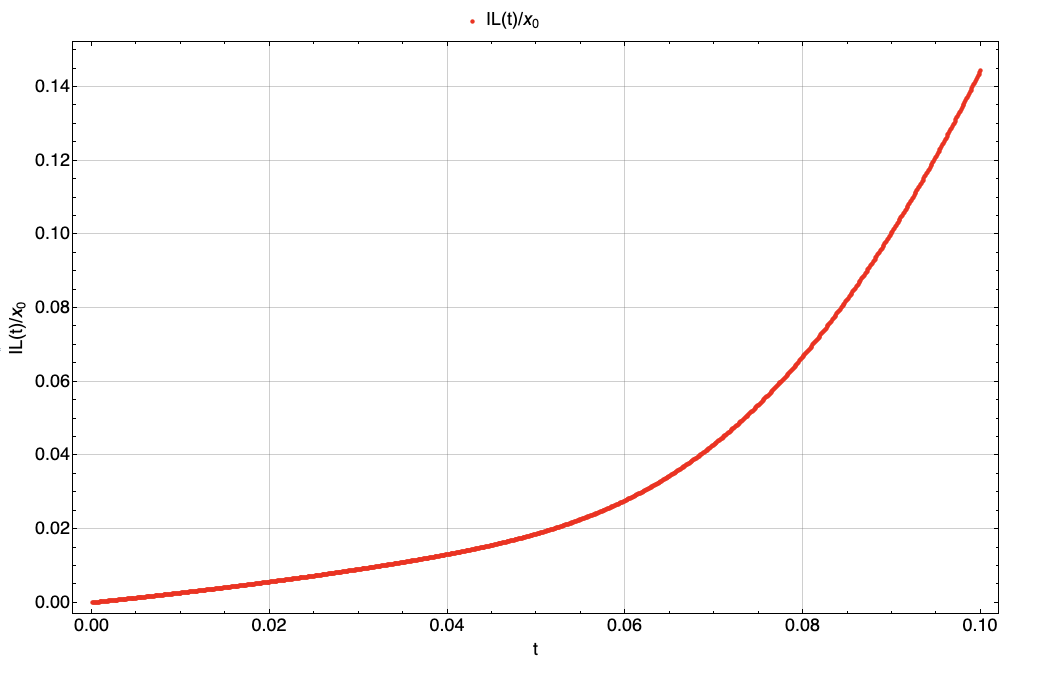}\caption{Expected IL as a function of time, $\langle IL (t) \rangle$, measured in units of $x_0$. For this plot we chose $\sigma_0=p_0=1$.}\label{fig:IL}
\end{figure}

In a first step, we analyze the expected IL as a function of time. It turns out that this quantity can readily be calculated from summing IL over all possible paths in price space:
\begin{eqnarray}
\langle {\rm{IL}}(t) \rangle &=& \int dp\frac{{\rm{IL}}(p)}{\sqrt{2\pi \sigma_0^2 t}}\exp{\left(-\frac{\left(p-p_0\right)^2}{2 \sigma_0^2 t}\right)} \nonumber \\ &=& \frac{x_0}{\sqrt{2\pi \sigma_0^2 t}}\int dp \left(1-\sqrt{\frac{p_0}{p}} \right)^2\exp{\left(-\frac{\left(p-p_0\right)^2}{2 \sigma_0^2 t}\right)} \nonumber \\ &=& \frac{x_0}{\sqrt{\pi}}\int dp \left(1-\sqrt{\frac{p_0}{p_0+\sqrt{2\sigma_0^2t}p}} \right)^2\exp{\left(-p^2\right)}\;.
\end{eqnarray}
We find that this integral does not extend to $-\infty$ but has to be cut off at $-p_0/\sqrt{2 \sigma_0^2 t}$. For practical purposes and short times $t\ll 2\sigma_0^2p_0^2$ we can expand the integrand to yield
\begin{eqnarray}
\langle {\rm{IL}}(t) \rangle \approx  \frac{x_0 \sigma_0^2}{4 p_0^2} t \int dp \frac{2 p^2 }{\sqrt{\pi}}\exp{\left(-p^2\right)}=\frac{x_0 \sigma_0^2}{4 p_0^2} t
\end{eqnarray}
which implies that the expected IL increases linearly with time (this integral can now be extended all the way to $-\infty$). A full numerical solution of the integral with its actual boundaries is shown in Fig.~\ref{fig:IL} but not important for our discussion. It just serves as a proof of validity of our expansion. 

The analytical formula captures the linear part well and could also be used to characterize the deviation if higher orders were takein into account. 

We now move to average LVR. It can be calculated from

\begin{eqnarray}
\langle {\rm{LVR}}(t) \rangle &=& \int dp \int_0^t dt' \, \rho(p,t') \frac{d {\rm{LVR}}(p)}{dt'} \nonumber \\ 
&=& \frac{x_0 \sigma_0 \sqrt{p_0}}{4\sqrt{2\pi }} \int dp \int_0^t dt' \frac{1}{\sqrt{t'}p^{5/2}}\exp{\left(-\frac{\left(p-p_0\right)^2}{2 \sigma_0^2 t'}\right)} \nonumber \\ 
&=& \frac{x_0 \sigma_0^2 \sqrt{p_0}}{4\sqrt{\pi }} \int dp \int_0^t dt' \frac{1}{(p_0+\sqrt{2\sigma_0^2 t'}p)^{5/2}}\exp{\left(-p^2\right)} \;.
\end{eqnarray}
We can expand the integrand to lowest order as before and get
\begin{eqnarray}
\langle {\rm{LVR}}(t) \rangle &\approx&  \frac{x_0 \sigma_0^2 }{4\sqrt{\pi }p_0^2} \int dp \int_0^t dt' \exp{\left(-p^2\right)} =\frac{x_0 \sigma_0^2 }{4p_0^2}t\;.
\end{eqnarray}
We thus conclude that we manged to show that $\langle {\rm{IL}}(t) \rangle=\langle {\rm{LVR}}(t) \rangle$, as we already observed from the numerical simulation. Furthermore, the numerical findings are in excellent agreement with the analytical predictions. The analytical prediction for the plots shown in Fig.~\ref{fig:MCILvsLVRa} and Fig.~\ref{fig:MCILvsLVRb} is $\langle {\rm{LVR}}(5000) \rangle=\langle {\rm{IL}}(5000) \rangle=0.00125$ for both averages.

\section{Distribution function of IL}\label{app:clt}

In this section we give a detailed derivation of the distribution function of IL. The starting point is the forumla for IL going from a starting price $p_0$ to a final price p given by
\begin{eqnarray}
{\rm{IL}}(p_0,p)=\frac{L}{\sqrt{p_0}}\left(1-\sqrt{\frac{p_0}{p}} \right)^2
\end{eqnarray}
We invert this expression and find the following solution:
\begin{eqnarray}
p({\rm{IL}})=\frac{p_0}{\left(p_0^{1/4}\sqrt{{\rm{IL}}/L}+1 \right)^2}\Theta(p_0-p)+\frac{p_0}{\left(1-p_0^{1/4}\sqrt{{\rm{IL}}/L} \right)^2}\Theta(p-p_0),
\end{eqnarray}
where ${\rm{IL}}=0$ corresponds to the point where $p=p_0$.
We find that the distribution function splits into two parts depending on whether $p>p_0$ or not:
\begin{eqnarray}
\bar{\rho}_{\rm{BM/GBM}}({\rm{IL}})&=&\frac{1}{\sqrt{{\rm{IL}}}}\frac{p_0^{5/4}}{\sqrt{L}} \frac{\rho_{\rm{BM/GBM}}\left(p_0/\left(p_0^{1/4}\sqrt{{\rm{IL}}/L}+1 \right)^2\right)}{\left(1+p_0^{1/4}\sqrt{{\rm{IL}}/L}\right)^3} \nonumber \\ &+&\frac{1}{\sqrt{{\rm{IL}}}}\frac{p_0^{5/4}}{\sqrt{L}} \frac{\rho_{\rm{BM/GBM}}\left(p_0/\left(1-p_0^{1/4}\sqrt{{\rm{IL}}/L} \right)^2\right)}{\left(1-p_0^{1/4}\sqrt{{\rm{IL}}/L}\right)^3}\theta \left( L/\sqrt{p_0}-{\rm{IL}}\right)\;.\nonumber \\
\end{eqnarray}
\end{document}